\documentclass[nofootinbib,aps,twocolumn,prl,superscriptaddress,nobibnotes]{revtex4-2}
\usepackage{graphicx}
\usepackage{physics}
\usepackage{dsfont}
\usepackage{amsfonts}
\usepackage{xfrac}
\usepackage{appendix}
\usepackage{hyperref}
\usepackage{enumitem}
\usepackage{subcaption}
\usepackage{bbding}
\usepackage{xcolor}
\usepackage{amssymb}
\usepackage{xr-hyper}
\usepackage{bibunits}
\usepackage{mathtools}

\makeatletter
\newcommand{\TocHideStart}{\let\TocSavedContentsline\contentsline\let\contentsline\@gobblefour}
\newcommand{\TocHideEnd}{\let\contentsline\TocSavedContentsline}
\makeatother

\newcommand{\C}{\mathds C}
\newcommand{\R}{\mathds R}
\newcommand{\E}{\mathds E}
\newcommand{\ME}{\text{ME}}
\newcommand{\CWE}{\text{CWE}}
\newcommand{\FTCWE}{\text{FTCWE}}
\newcommand{\CN}{\mathcal{CN}}
\newcommand{\POVM}{\text{POVM}}
\newcommand{\dn}{\downarrow}

\usepackage{amsthm}
\theoremstyle{plain}
\newtheorem{thm}{Theorem}
\newtheorem{prop}[thm]{Proposition}
\newtheorem{lemma}[thm]{Lemma}
\newtheorem{cor}[thm]{Corollary}

\theoremstyle{definition}
\newtheorem{defn}[thm]{Definition}

\theoremstyle{remark}
\newtheorem{rem}[thm]{Remark}

\begin{document}

\title{Maximum-Entropy Ensemble of Quantum Subsystems with a Fixed Average State}

\author{Charlie Shakeshaft}
\thanks{Contact author: charlie.shakeshaft@postgrad.manchester.ac.uk}
\affiliation{Department of Physics \& Astronomy, University of Manchester, Manchester M13 9PL, United Kingdom}

\author{Christopher David White}
\affiliation{Center for Computational Materials Science, U.S. Naval Research Laboratory, Washington, D.C. 20375, USA}

\author{Harry J. D. Miller}
\affiliation{Department of Physics \& Astronomy, University of Manchester, Manchester M13 9PL, United Kingdom}

\date{\today}

\begin{abstract}
    The reduced density matrices of Haar-random bipartite states form a fixed-trace version of the \textit{Wishart ensemble} of random matrices, with an asymptotic eigenvalue density given by the Marchenko--Pastur (MP) law. We establish a similar universal form for the maximum-entropy distribution of mixed states subject to a fixed ensemble average, arguing that its eigenvalue density converges asymptotically to that of the \textit{correlated} Wishart ensemble (CWE). The density is thus obtained by a free multiplicative convolution formula, which we apply to the transverse-field Ising model (TFIM) with a thermal average state to derive a high-temperature leading-order correction to the MP law. 
    We then consider measurement by a random POVM on the purification of the mixed reference state.
    We show that the resulting mixed-state ensemble is asymptotically equivalent to the CWE, and provide confirming numerical evidence. We return to the TFIM example to compute the entanglement of the system with the ancilla after measurement, and give finite-temperature Page curves for the average entanglement entropies.
\end{abstract}
\maketitle
\textit{Introduction}.---Random matrix theory (RMT) is a central tool in understanding both the emergence of universal spectral features in quantum chaotic systems~\cite{QChaosETH,QSignaturesChaosHaake} and the typical properties of random mixed states~\cite{StatsRDM,GenRDM,GQS,StatMechFixedEnergy,EntropyBasedQStates,RMInQIT,DevelopmentsRMT}. Though there is no unique probability measure on the set of mixed states, a natural ensemble is induced by tracing out either subsystem of a global Haar-random bipartite state~\cite{InducedMeasures,GenRDM}. The statistical properties of mixed states under this induced measure, such as the moments of purity~\cite{Lubkin1978,InducedMeasures,Giraud} and entanglement entropy~\cite{AvgEntropySubsystem,PageConjectureProof,PageProofSimple,EntanglementVarianceConjecture,EntanglementVarianceProof,EntanglementSkewness}, have been studied extensively, and the distribution is equivalent to the \textit{Wishart ensemble} of sample covariance matrices~\cite{Wishart1928} up to the unit trace constraint~\cite{InducedMeasures,AsymptoticsRDM}. Moreover, due to the concentration of the traces of the Wishart matrices around their expectation value in the limit of large dimensions, both the induced and Wishart ensembles share a single-point eigenvalue density given by the \textit{Marchenko--Pastur} (MP) law~\cite{MarchenkoPastur,AsymptoticsRDM}.

However, global Haar randomness is of limited applicability in realistic systems due to finite-temperature constraints imposed by conservation laws. While universal ensembles have been established for pure states with a fixed ``background'' density matrix~\cite{ScroogeManyBody,NatureIsStingy,DeepThermChargeCons,MaxEntDeepTherm,MaxEntWavefunctions}, as well as for mixed states under error channels at infinite temperature~\cite{MixedStateDeepThermalisation,MSDTErrors}, the incorporation of constraints into the treatment of random mixed states remains an open problem. In this work, we bridge the gap by constructing a maximum-entropy ensemble of quantum subsystems subject to a fixed average state. We derive an expression for the partition function of the ensemble, which can be used to compute expectation values and fluctuations. In the limit of large dimensions, we show that the eigenvalue density of this maximum-entropy distribution converges to that of the \textit{correlated Wishart ensemble} (CWE): an extension of the Wishart ensemble to the case of nontrivial statistical correlations within the underlying data~\cite{CorrelatedWishartEnsembles,SpectralMomentsCWE,SignalNoiseCWE,NoiseDressingCWE,DistributionForCWE,ExtremeEigenvaluesCWE}. Exploiting this asymptotic equivalence, we apply tools from free probability theory~\cite{Voiculescu,FreeRandomVariables,FreeProbRMT} to express the eigenvalue density of the constrained states in terms of a \textit{free multiplicative convolution} between the unconstrained MP law and the spectral measure over the reference state. For the transverse-field Ising model (TFIM) with a thermal reference state, we evaluate this convolution via Voiculescu's S-transform~\cite{Voiculescu}, arriving at an implicit expression for the level density that reduces to a leading-order correction to the MP law in the high-temperature regime. 

Finally, we show that the mixed-state ensemble generated by a random local measurement on the canonical purification of the chosen reference state has CWE-distributed eigenvalues asymptotically, providing supporting numerical evidence. We compute the Page curves~\cite{AvgEntropySubsystem} for the average entanglement entropy of the subsystem distribution generated by the random measurement, and find that these are both suppressed due to the thermal bias and asymmetric about the infinite-temperature midpoint.

\textit{Maximum-entropy ensemble}.---Consider a Haar-random bipartite state $\ket{\psi}\in \mathcal{H}_S\otimes\mathcal{H}_E$, with $\dim(\mathcal{H}_S) = d$ and $\dim(\mathcal{H}_E) = K\geq d$. Tracing out the environment $E$, the reduced density matrices $\rho = \Tr_E \ketbra{\psi}{\psi}$ are distributed according to the \textit{induced measure} on the set $\mathcal{S}(\mathcal{H}_S)$ of mixed states, with probability density~\cite{GQS,GenRDM}
\begin{equation}
\label{InducedMeasure}
    P_K(\rho) \propto \left(\det \rho\right)^{K-d} \Theta(\rho) \, \delta(\Tr \rho - 1) ,
\end{equation}
where the $\Theta$ and $\delta$ functions enforce positive-semidefiniteness and unit trace respectively. Note that for $K<d$, the states are no longer full-rank and the base measure is supported on a lower dimensional manifold. Though we restrict to $K\geq d$ in our entropy maximisation, our later large-dimensional equivalence between the CWE and its fixed-trace version holds for any aspect ratio $d/K$. Under the diagonalisation $\rho = U D U^\dagger$, where $D = \text{diag}(r_1, \dots, r_d)$ contains the eigenvalues $\{r_i\}_{i=1,\dots , d}$ of $\rho$ and $U$ is a unitary with the eigenvectors as its columns, the measure $d\mu_K(\rho) = P_K(\rho)d\rho$ factorises into a contribution $d\mu_K(\vec{r})$ over the eigenvalues and the unitary Haar measure $d\mu_{\text{Haar}}(U)$ over the eigenvectors. In the joint limit $d, K \to \infty$ such that $d/K \to c \in (0,\infty)$~\cite{NonlinearShrinkage}, the empirical distribution of the rescaled eigenvalue $x = d r_i$ converges to the MP distribution~\cite{MarchenkoPastur,AsymptoticsRDM}
\begin{align}
\label{MarchenkoPastur}
    \pi_c(x)&= \left(1-\frac{1}{c}\right)\delta(x)\, \mathds{1}_{\{c>1\}}\nonumber\\
    &+\frac{\sqrt{(x_+-x)(x-x_-)}}{2\pi c x}\, \mathds{1}_{\{x\in [x_-,x_+]\}},
\end{align}
with edges $x_\pm = (1\pm\sqrt{c})^2$. A further celebrated result is the formula first conjectured by Page~\cite{AvgEntropySubsystem} for the average von Neumann entropy of the random states, a simple proof of which was subsequently given using the properties of the Wishart ensemble~\cite{PageProofSimple}. 

We investigate how fixing the expectation value $\langle\rho \rangle \coloneq \rho_0$ impacts these spectral properties. Following Jaynes' principle~\cite{JaynesI,JaynesII}, we maximise the relative entropy\footnote{defined here as the \textit{negative} Kullback-Leibler divergence between the distributions. If defined with the opposite sign, our solution is obtained by minimisation.} $-\int_{\mathcal{S}(\mathcal{H}_S)} P(\rho) \ln \left[P(\rho)/P_K(\rho)\right] d\rho$ subject to the constraint to obtain the probability density~\cite[][Sec.~\ref{app:Max-Ent-measure}]{Supplementary}
\begin{equation}
\label{MaxEntDensity}
    P_{\text{ME}}(\rho) = \frac{1}{Z_K(M)}\, e^{-\Tr M\rho}\, P_K(\rho) ,
\end{equation}
where $M$ is a $d\times d$ Hermitian matrix of Lagrange multipliers enforcing the constraint, and the normalisation factor $Z_K(M)$ is the partition function of the ensemble. Writing the base measure $d\mu_K(\rho)$ in the eigenvalue-eigenvector product form, the latter is given by the integral
\begin{equation}
    Z_K(M)= \int_{\mathcal{S}(\mathcal{H}_S)} d\mu_K(\vec{r})\times d\mu_{\text{Haar}}(U)\ e^{-\Tr M U D U^\dagger}.
\end{equation}
We evaluate this analytically in~\cite[][Sec.~\ref{app:partition-function}]{Supplementary}, generalising the \textit{Hilbert-Schmidt} ($d = K$) partition function derived in~\cite{StatMechFixedEnergy} to the case of an arbitrary environment. Analogues of the canonical statistical-mechanical functions follow as derivatives in the standard way, and the matrix $M$ itself is determined by the set of $d^2-1$ independent constraint equations
\begin{align}
\label{ConstraintEqns}
    -\frac{\partial \ln Z_K(M)}{\partial M}&= \rho_0.
\end{align}
Note that the solutions are defined up to the additive gauge $M\mapsto M + \alpha I$, $\alpha\in\mathbb{R}$. We argue later that, with the gauge fixed such that $\Tr M^{-1} = 1/K$, $M/K$ approaches $\rho_0^{-1}$ as $K\to\infty$.

In the physical situation we consider in both our TFIM example and the single-qubit case in the \hyperref[EndMatter]{End Matter} is $\rho_0$ fixed to the thermal state of the subsystem Hamiltonian. Fluctuations around equilibrium are then encoded in higher moments of the ensemble, and the deviations from the MP and Page predictions represent finite-temperature corrections to the spectra of the mixed states.

\textit{Asymptotic spectral properties}.---Many properties of a density matrix---the von Neumann entropy, for example---are controlled by its spectrum, and in RMT one is often interested in whether the empirical spectral distribution of an ensemble converges asymptotically to a deterministic density. Here, we argue that the spectral properties of the maximum-entropy ensemble become equivalent in the limit of a large Hilbert space to those of the fixed-trace version of the CWE. 

Correlated Wishart matrices are of the form~\cite{CorrelatedWishartEnsembles}
\begin{equation}
\label{CWEMatrix}
    W = \frac{1}{K}\Sigma^{1/2}XX^\dagger \Sigma^{1/2},
\end{equation}
where $X$ is a $d\times K$ \textit{Ginibre matrix} with complex Gaussian-distributed entries $X_{ij}\sim \mathcal{C}\mathcal{N}(0,1)$, and $\Sigma \succ 0$ is a positive-definite matrix encoding the correlations between the elements. The probability density of the CWE is given by~\cite{CorrelatedWishartEnsembles,SpectralMomentsCWE}
\begin{equation}
\label{CWEDensity}
    P_{\CWE}(W) =K^{dK}\frac{\det(W)^{K-d}}{\Gamma_d(K) \det(\Sigma)^{K}} \, e^{-K\Tr \Sigma^{-1}W} ,
\end{equation}
where $\Gamma_d(K) = \pi^{d(d-1)/2}\prod_{j=1}^d \Gamma(K-j+1)$.
The CWE has average matrix $\langle W\rangle=\frac{1}{K}\Sigma^{1/2}\langle XX^\dagger\rangle\Sigma^{1/2}=\Sigma$. 
To obtain the density of the fixed-trace CWE (FTCWE), we define the positive-semidefinite, unit-trace matrix $\rho$ by decomposing $W = t\rho$, with $t = \Tr W > 0$. Marginalising Eq.~\eqref{CWEDensity} over this random trace $t$ yields~\cite[][Sec.~\ref{app:FTCWE-density}]{Supplementary}
\begin{equation}
\label{FTCWEDensity}
    P_{\text{FT}}(\rho) = \frac{\Gamma(dK)}{\Gamma_d(K)\det(\Sigma)^K}\frac{\det(\rho)^{K-d}}{\left(\Tr \Sigma^{-1}\rho\right)^{dK}}\, \Theta(\rho)\, \delta\left(\Tr \rho-1\right).
\end{equation}
In what follows, we fix the gauge of $M$ such that $\Tr M^{-1}=1/K$.

The main result of this section is two-fold. We show that (i) the maximum-entropy ensemble with $M$ such that $\langle\rho\rangle=\rho_0$ and the FTCWE with covariance matrix $\Sigma = \rho_0$ become equivalent in the limit of large $K$, giving a non-rigorous argument that $M \approx K\rho_0^{-1}$ in this limit. 

 We then show that (ii) due to the large-$K$ concentration of the trace variable $t$, the empirical spectral distributions of the FTCWE and the CWE converge in probability in the joint $d,K$ limit. Hence, the level density of our ensemble is obtained by evaluating the \textit{free multiplicative convolution}
\begin{equation}
\label{FreeConvolution}
    \mu_{\text{CWE}} = \mu_{\rho_0}\boxtimes \mu_{c}
\end{equation}
between the limiting spectral measure $\mu_{\rho_0}$ of the (rescaled) reference state $d\rho_0$ and the MP measure $\mu_c$~\cite{Voiculescu,FreeProbRMT,AnisotropicLocalLaws}.

To sketch our physical argument for (i), we first decompose both the maximum-entropy and FTCWE density matrices in terms of normalised, non-Hermitian random matrices defining distributions on the unit sphere $S^{2dK-1}\subset \mathbb{C}^{d\times K}$. Let $\rho_{\text{ME}} = \omega\omega^\dagger$ be a density matrix distributed according to Eq.~\eqref{MaxEntDensity}, with Lagrange multiplier $M =K \Sigma^{-1}$. The $d\times K$ matrix $\omega$ then has the PDF
\begin{equation}
    p_{\text{ME}}(\omega) \propto e^{-\Tr M\omega\omega^\dagger} = e^{-dK q(\omega)},
\end{equation}
where we have defined $q(\omega) = \frac{1}{dK}\Tr M\omega\omega^\dagger$. We identify this as the \textit{complex Bingham distribution} on the sphere~\cite{Bingham,ComplexBinghamDist}. Likewise, let $\rho_{\text{FT}} = \varphi\varphi^\dagger$ be FTCWE matrices distributed according to Eq.~\eqref{FTCWEDensity}. The PDF of the matrix $\varphi$ is then
\begin{equation}
    p_{\text{FT}}(\varphi) \propto \left(\Tr \Sigma^{-1}\varphi\varphi^\dagger\right)^{-dK} \propto q(\varphi)^{-dK} .
\end{equation}
As we argue in~\cite[][Sec.~\ref{app:Max-Ent-to-FTCWE}]{Supplementary}, both densities on the sphere approach the same limit at large $K$,
\begin{equation}
    p_{q}(\varphi) \propto \delta\left(1-q(\varphi)\right),
\end{equation}
i.e.\ a ``microcanonical ensemble'' with respect to the random variable $q$.
Indeed in \cite[][Sec.~\ref{sss:integrability}]{Supplementary}
we rigorously show a stronger result:
that both distributions approach a ``generalised microcanonical ensemble''
in which the weight on each eigenspace of $\Sigma$ is fixed.

Returning to the CWE with $\Sigma=\rho_0$ ($\Tr\Sigma=1$), result (ii) relies on the Bernstein inequality~\cite{HighDimensionalProbability}
\begin{equation}
    P\left(\abs{\Tr W -1}\geq t\right)\leq 2\exp\left[-\kappa\min\left(\frac{9Kt^2}{32\Tr \rho_0^2},\frac{3Kt}{4\norm{\rho_0}_{\text{op}}}\right)\right],
\end{equation}
where $\kappa$ is an absolute constant. We prove this in~\cite[][Sec.~\ref{app:concentration-normalization}]{Supplementary}. In the Gaussian core of the distribution of $\Tr W$, this inequality bounds the fluctuations in the variable according to the purity of the reference state via $K^{-1/2}\sqrt{\Tr \rho_0^2} = K^{-1/2}\,e^{-S_2(\rho_0)/2}$, where $S_p(\rho_0)$ denotes the R\'enyi entropy of order $p$. Fluctuations in the exponential tail are instead bounded by the operator norm $K^{-1}\norm{\rho_0}_{\text{op}}=K^{-1}e^{-S_\infty(\rho_0)}$. 

It follows from the bounds that, for $W$ distributed according to the CWE with covariance $\rho_0$ and $\rho = W/\Tr W$ distributed according to the corresponding FTCWE, the empirical spectral distributions of $\rho$ and $W$ converge in probability. Hence, the asymptotic formula~\eqref{FreeConvolution} applies to the maximum-entropy ensemble. A more detailed proof is provided in the Supplemental Material~\cite[][Sec.~\ref{app:secIII}]{Supplementary}.

\begin{figure}[h]
    \centering
    \includegraphics[width=0.9\linewidth]{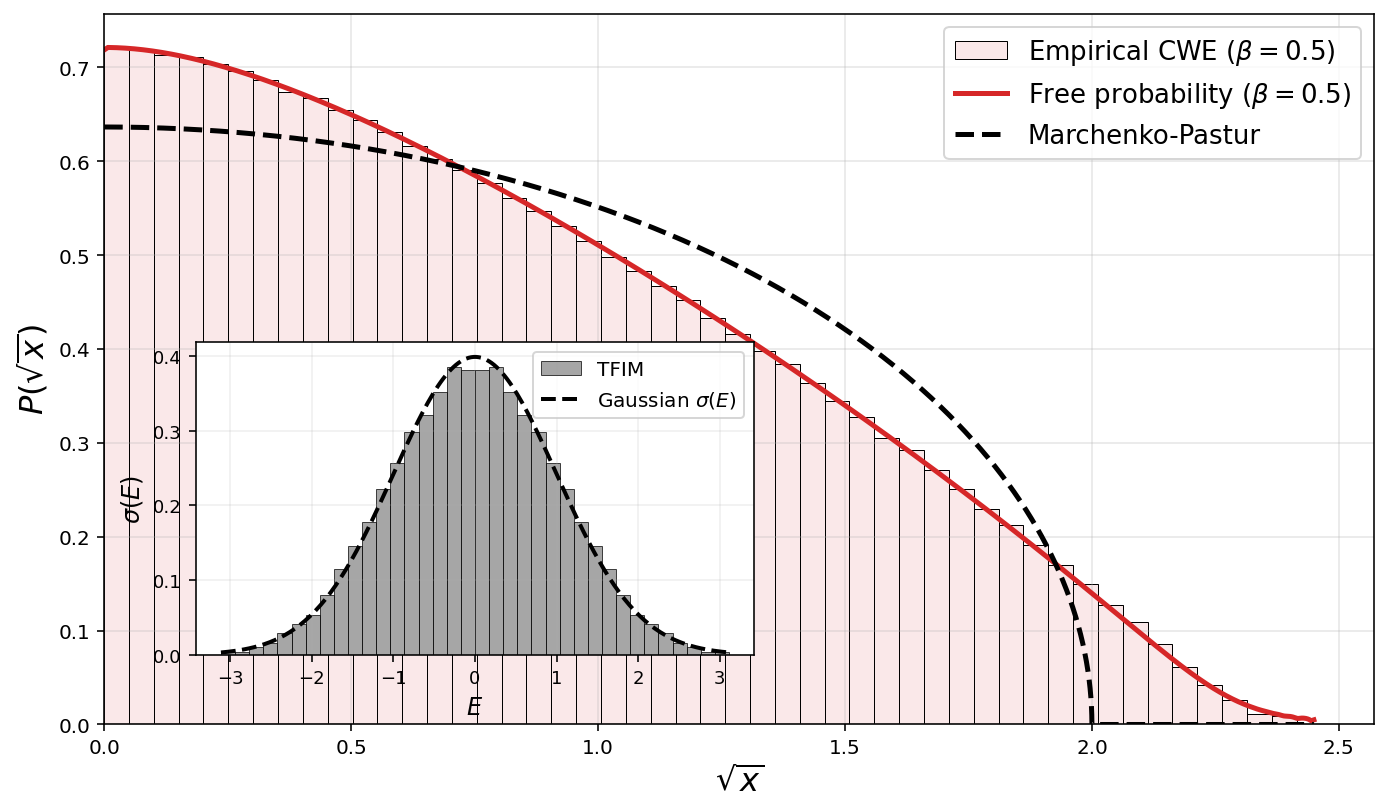}
    \caption{\small CWE level density in terms of $\sqrt{x} = \sqrt{d\lambda}$, $\lambda$ the eigenvalues of the CWE matrices, for $\rho_0$ the thermal state of the TFIM Hamiltonian (at $\beta = 0.5$, $c = 1$, $J = h$ with $H$ rescaled so that $\sigma_0 = 1$). The histogram shows the empirical distribution (with $\rho_0$ computed from the exact spectrum of $H$ at $N_d = 12$), while the solid line shows the free probability prediction using a Gaussian density of states. \textit{Inset}: Empirical density of the standardised energy $E/\sigma_0$ of the $12$-qubit TFIM ($\sigma_0^2=\Tr H^2/d$) compared to the Gaussian approximation.}
    \label{CWELevelDensity}
\end{figure}

We now compute the density for the example of the 1D TFIM with open boundary conditions. For $N_d$ qubits (with $d = 2^{N_d}$), the Hamiltonian reads
\begin{equation}
    H = -J \sum_{\langle i,j\rangle}\sigma_z^{(i)}\sigma_z^{(j)}-h\sum_{i=1}^{N_d}\sigma_x^{(i)},
\end{equation}
where $\langle i,j\rangle$ denotes nearest-neighbour pairs, and $J$ and $h$ are the respective strengths of the spin-spin and external field terms. We fix $\rho_0$ to the thermal state at inverse temperature $\beta$, and introduce an auxiliary environment of dimension $K$. For sufficiently large $N_d$, the energy density $\sigma(E)$ is well-approximated by a Gaussian with variance $\sigma_0^2 = \Tr H^2/d$ (see Fig.~\ref{CWELevelDensity}). We transform this to a density $\pi_{\rho_0}(x)=\sigma(E)\abs{\frac{dE}{dx}}$ over the rescaled eigenvalue $x = de^{-\beta E}/Z$ of the thermal state in~\cite[][Sec.~\ref{app:S-transform}]{Supplementary}, obtaining a log-normal expression. We then evaluate the constrained measure $\mu_{\CWE}$ using the definition of the free multiplicative convolution $\boxtimes$ in terms of the Voiculescu S-transform. This states that, for two asymptotically free random-matrix ensembles with limiting spectral measures $\mu$ and $\nu$ respectively, the convolution $\mu \boxtimes \nu$ is the unique probability measure for which the following relation holds in the S-transform domain~\cite{SpecDensGeneralisedWishart,FreeRandomVariables}:
\begin{equation}
    S_{\mu \boxtimes \nu}(z) = S_\mu(z) S_\nu(z),
\end{equation}
where $z$ is a complex variable. Conveniently, the S-transform of the MP measure $\mu_c$ is a standard result~\cite{RMTWireless}
\begin{equation}
\label{STransformMP}
    S_c(z) = \frac{1}{1+cz},
\end{equation}
while that of the thermal measure with density $\pi_{\rho_0}(x)$ is expressed implicitly as the integral~\cite[][Sec.~\ref{app:S-transform}]{Supplementary}
\begin{equation}
\label{STransformThermal}
    S_{\rho_0}(z)=(1+z)\left[\int_{\text{supp}[\sigma(E)]} \frac{\sigma(E) \, dE}{U(\beta) e^{\beta E}-\frac{z}{1+z}S_{\rho_0}(z)}\right]^{-1},
\end{equation}
where $U(\beta) = \int_{\text{supp}[\sigma(E)]}\sigma(E)\, e^{-\beta E}dE$ denotes the two-sided Laplace transform of the energy density. From these expressions, we obtain the level density by numerically inverting the product $S_c(z)S_{\rho_0}(z)$. Fig.~\ref{CWELevelDensity} shows an example for the Hilbert-Schmidt measure ($c=1$), where the infinite-temperature density is quarter-circular in the rescaled variable $\sqrt{x}$~\cite{GQS,GenRDM}.

In~\cite[][Sec.~\ref{app:Gaussian-spec-dens}]{Supplementary}, we expand the log-normal density $\pi_{\rho_0}(x)$ to leading order in $\beta\sigma_0$ to arrive at a high-temperature perturbative correction to the continuous part $\pi_\text{cont}(x)$ of the MP distribution:
\begin{equation}
\label{MPCorrection}
    \pi_\text{cont}(x) \approx \frac{\sqrt{4(c+\beta^2\sigma_0^2x)-(1+c-x)^2}}{2\pi x(c+\beta^2\sigma_0^2x)}.
\end{equation}
We provide the analogous expression for a Hamiltonian sampled at random from the Gaussian unitary ensemble in the Supplemental Material~\cite[][Sec.~\ref{app:GUE}]{Supplementary}.

\begin{figure}[t]
    \centering
    \includegraphics[width=0.9\linewidth]{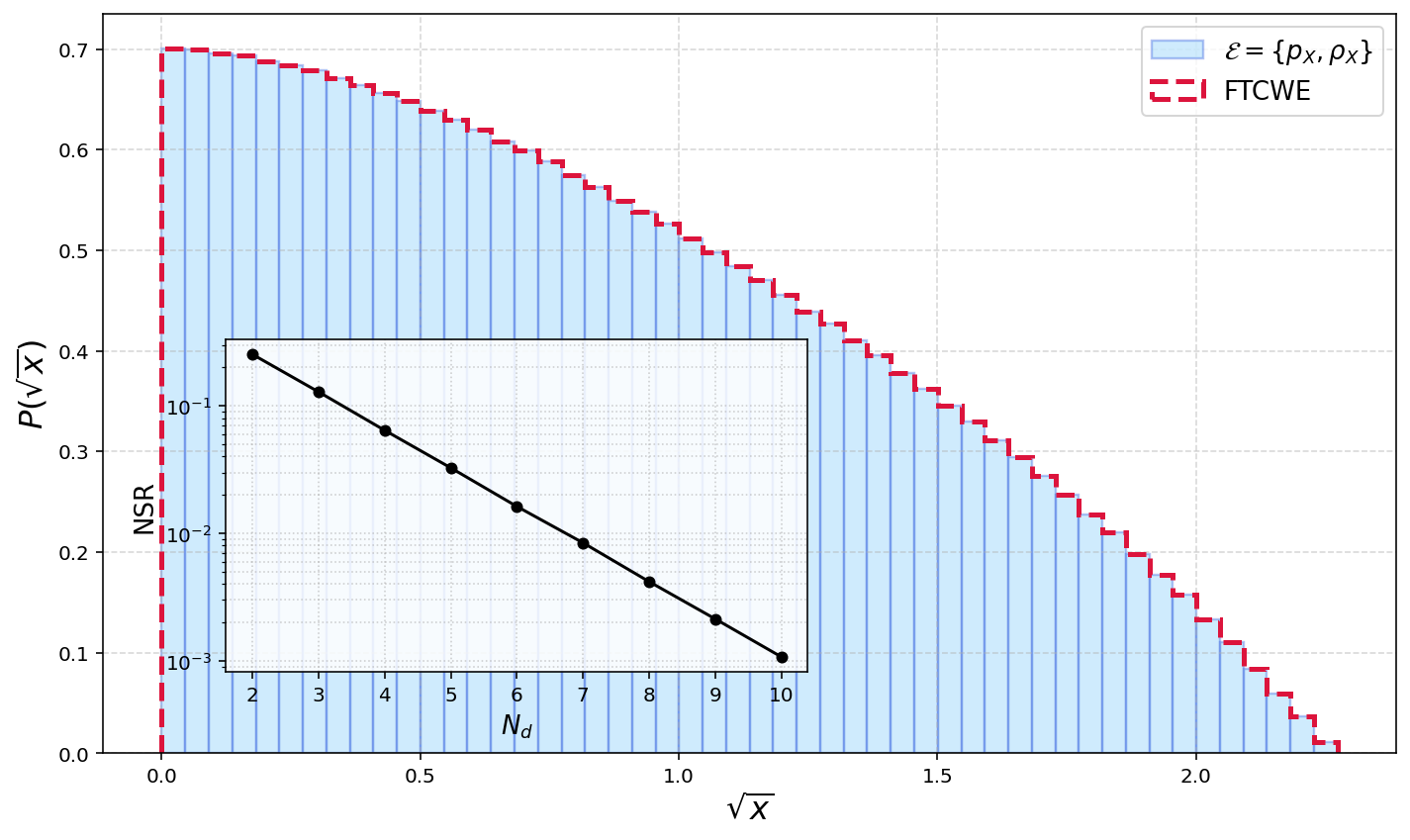}
    \caption{\small Empirical distribution of the rescaled eigenvalue $\sqrt{x}=\sqrt{d\lambda}$ for the ensemble $\mathcal{E}$ generated by the random POVM compared to that of the FTCWE. We use the $N_d = 10$ qubit TFIM, with $c = 1$ and $\beta = 0.1$. \textit{Inset}: exponential decay of the noise-to-signal ratio (NSR) $\frac{\sqrt{\text{var}(\Tr W_X)}}{\langle \Tr W_X \rangle}$ as a function of system size.}
    \label{Fig:POVM-vs-FTCWE}
\end{figure}

\textit{Random POVM}.---We propose the following measurement scheme as a physical mechanism for the emergence of the CWE level density. Let the reference state on a $d$-dimensional Hilbert space $\mathcal{H}_A$ be $\rho_0 = \sum_{i=1}^d \lambda_i\ketbra{i}{i}$. We introduce the auxiliary space $\mathcal{H}_B$ of the same dimension, and consider the canonical purification
\begin{equation}
\label{Purification}
    \ket{\psi}_{AB} = \sum_{i=1}^d \sqrt{\lambda_i}\ket{i}_A\otimes \ket{i}_B  .
\end{equation}
We then perform a random measurement on subsystem $B$, modelled by the Kraus operator $M_X = \frac{1}{\sqrt{K}}X^T$ (with $X$ again a $d \times K$ Ginibre matrix). The operation maps $\mathcal{H}_B \to \mathcal{H}_C \cong \mathbb{C}^K$, and we verify that the (continuously indexed) set of effects $\{ M_X^\dagger M_X \}$ obeys the POVM condition
\begin{equation}
    \int d\mu(X) \, M_X^\dagger M_X = I_d,
\end{equation}
where $d\mu(X) = \pi^{-dK}\, e^{-\Tr XX^\dagger} dX$ is the Gaussian measure over the complex Ginibre ensemble. Conditioned on the specific outcome $X$, the (unnormalised) state of the system after measurement is
\begin{equation}
    \tilde{\ket{\psi}}_{AC} = \left( I_d \otimes M_X \right)\ket{\psi}_{AB},
\end{equation} 
with the subsystem $A$ described by the positive-semidefinite matrix
\begin{equation}
    W_X = \Tr_C \tilde{\ket{\psi}}\tilde{\bra{\psi}}=\frac{1}{K}\, \rho_0^{1/2}XX^\dagger\rho_0^{1/2}.
\end{equation}
We prove the second equality--- i.e., that tracing out $C$ on the post-measurement state yields the correlated Wishart form--- in~\cite[][Sec.~\ref{app:random-POVM}]{Supplementary}. The physical states are then simply $\rho_X \coloneq  W_X / \Tr W_X$, each weighted by the Born-rule probability density $p_X = \Tr W_X$. The resulting ensemble $\mathcal{E} = \{p_X, \rho_X\}$ has first moment $\rho_0$, and probability density
\begin{equation}\label{eq:povm-pdf}
    P_{\mathcal{E}}(\rho) \propto \frac{\det(\rho)^{K-d}}{\left(\Tr \rho_0^{-1}\rho\right)^{dK+1}}\, \Theta(\rho)\, \delta(\Tr \rho -1),
\end{equation}
which, comparing to Eq.~\eqref{FTCWEDensity}, differs from the FTCWE with covariance $\rho_0$ by a factor of $(\Tr \rho_0^{-1}\rho)^{-1}$.
In \cite[][Sec.~\ref{sss:integrability}]{Supplementary} we rigorously show that the POVM ensemble nonetheless approaches the FTCWE,
and in Fig.~\ref{Fig:POVM-vs-FTCWE} we verify numerically that the Born weightings have a negligible effect on the asymptotic level density; indeed, we observe exponential concentration with respect to $N_d$ for the variable $\Tr W_X$.

Note that $\mathcal{E}$ is spectrally equivalent to the distribution recently shown to emerge under scrambling dynamics interspersed with local errors~\cite{MSDTErrors}, given that the reference state in the former construction matches the Choi state of the error channel in the latter. We also comment that for $K=1$, $\mathcal{E}$ reduces to the \textit{Scrooge ensemble}: the distribution of pure states with first moment $\rho_0$ that minimises the accessible information~\cite{JoszaScrooge,NatureIsStingy,ScroogeManyBody,MaxEntDeepTherm}.

\begin{figure}[h]
    \centering
    \includegraphics[width=0.9\linewidth]{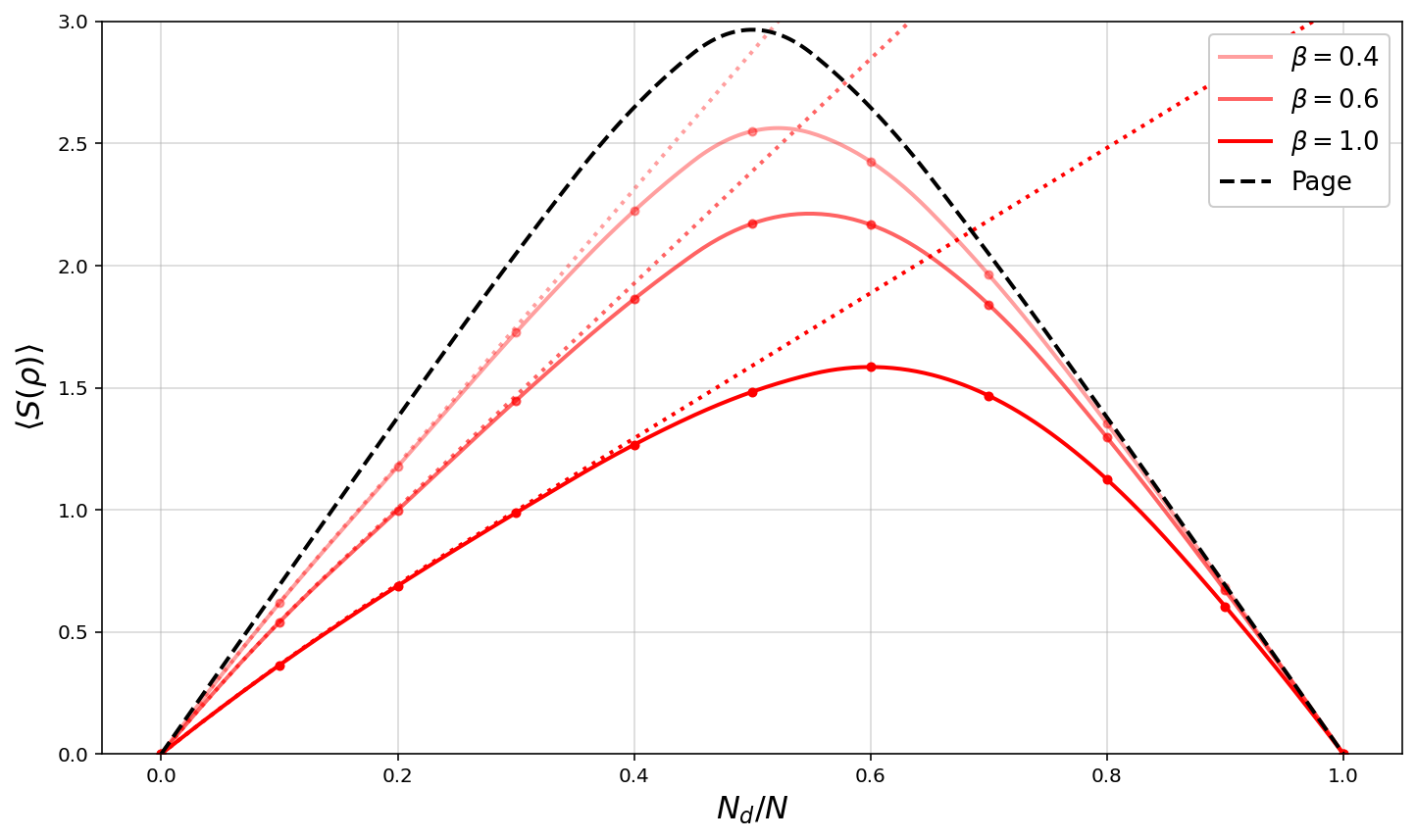}
    \caption{ \small Page curves for the POVM-generated ensemble $\mathcal{E}$, with $\rho_0$ the thermal state of the $N_d$-qubit TFIM. We fix $N = N_d + N_K = 10$, defining $N_K$ such that $K = 2^{N_K}$. The dotted lines show the (extensive) von Neumann entropy of the thermal state at each temperature $\beta^{-1}$.}
    \label{PageCurve}
\end{figure}

We now return to the TFIM example with $\rho_0$ thermal. The purification~\eqref{Purification} is then a thermofield double state, and the eigenvalues of the measured ensemble $\mathcal{E}$ are asymptotically distributed according to Eq.~\eqref{FreeConvolution}. In Fig.~\ref{PageCurve}, we vary the system-environment partitioning to compute Page curves of the $\mathcal{E}$-averaged von Neumann entropy of the mixed states, i.e. the average entanglement entropy of the global pure-state ensemble generated by the measurement. At the Hilbert space dimension we consider ($dK = 2^{10}$), the curves generated by this direct sampling of $\mathcal{E}$ agree with those obtained via free probability theory (see Fig.~\ref{Page-curve-POVM-vs-free-prob} in~\cite[][Sec.~\ref{app:random-POVM}]{Supplementary}).

We observe that the thermal bias on the subsystem reduces the entropy of typical states and breaks the symmetry about the Hilbert-Schmidt midpoint. Indeed, our environment remains maximally mixed on average; if instead we were to impose the constraint on both, we anticipate that the relevant asymptotic form would be the \textit{doubly correlated} Wishart ensemble~\cite{SpectralMomentsCWE,ApplyingFreeRVs}, and that the Page curves would remain symmetric under $d \leftrightarrow K$. The proximity of $\langle S(\rho)\rangle$ to the extensive quantity $S(\rho_0)$ in the regime $d\ll K$ again indicates concentration around the thermal state, with subsystems in the opposite regime being far-from-thermal and typically high purity.

\textit{Discussion}.---In this work, we have introduced the maximum entropy distribution of subsystem states with a fixed ensemble average $\rho_0$, representing the Bayesian ``best guess'' given the prior of global Haar randomness and the updated information $\langle\rho\rangle = \rho_0$. In the context of \textit{deep thermalisation}, similar maximum-entropy principles underpin the distributions that emerge at long times when the so-called projected ensembles of pure states conditioned on measurement outcomes on their complement evolve under generic dynamics~\cite{MaxEntDeepTherm,NatureIsStingy,DeepThermChargeCons}.

Our central result is the asymptotic convergence of the eigenvalue density of this information-theoretic ensemble to that of the CWE, alongside the emergence of the same CWE statistics under a random POVM on the purified reference state. This connects the CWE to the
physical setting of noisy or unsharp measurement.

Our results open up a number of avenues for future work. First, the maximum-entropy formalism is readily generalisable to the problem of $n$ linear constraints on the observables $\langle \Tr \mathcal{O}_i\rho\rangle$, with $i = 1, \dots, n$ (see~\cite{ConjugateVariables} for such a construction over classical probability distributions, and~\cite{BrodyInfoContent} for random pure states). In quantum thermodynamics, one may apply the formalism to compute the work extracted from a typical constrained state in a many-body system, or to characterise the performance of such states as quantum batteries using the ergotropy. Indeed, it may be possible to establish asymptotic concentration behaviour for the ergotropy, analogously to the infinite-temperature result~\cite{ConcOfErg}. A further natural extension is to verify that constraining both the system and environment corresponds to the doubly correlated case in the limit, with the goal of establishing the universal form for finite-temperature mixed-state deep thermalisation that reduces to the Scrooge ensemble at the pure-state level~\cite{MSDTErrors}.

Aside from the induced/Wishart ensemble, we raise the question of whether universal RMT behaviour emerges under similar constraints on other natural quantum ensembles, such as those induced by the Bures~\cite{GQS,BuresVol,Petz,BHAvgEntropy,RandomBuresPurity} or Bogoliubov-Kubo-Mori (BKM)~\cite{Balian,EntropyBasedQStates} metrics on the mixed-state manifold.

\textit{Acknowledgements}.---HJDM and CS acknowledge funding from a Royal Society Research Fellowship (URF/R1/231394). CDW is grateful to Christopher David White for helpful conversations in the context of a related project. This work was performed while CDW held an NRC Research Associateship award at the United States Naval Research Laboratory.
Claude 4.x and 5 and GPT 5.6 Sol suggested many proof strategies and individual steps,
especially in Sec.~II of the Supplemental Material;
the proofs as presented were all ultimately written by the human authors.
GPT 5.6 Sol and other models were used to assist with the numerical inverse S-transform code and to check correctness of the final manuscript.

\textit{Data availability}.---The code used to generate all numerical results and figures is available at~\cite{Code-For-Figures}.

\addtocontents{toc}{\protect\setcounter{tocdepth}{-1}}

\addtocontents{toc}{\protect\TocHideStart}

\bibliographystyle{apsrev4-2}
\bibliography{bibliography-CWE-paper}
\section*{End matter: Single qubit subsystem}
\label{EndMatter}
Here, we apply the maximum-entropy construction to the example of a single-qubit subsystem under the local Hamiltonian $H$, with $\rho_0$ fixed to the thermal state at inverse temperature $\beta$. Linear statistics of an arbitrary observable $\mathcal{O}$, such as the (ensemble) expectation value $\langle \Tr \mathcal{O}\rho\rangle$, match the equilibrium predictions $\Tr \mathcal{O}\rho_0$ by construction, while higher cumulants $\kappa_n$ (encoding fluctuations in the variable $\Tr \mathcal{O}\rho$) are computed using the perturbation formula~\cite[][Sec.~\ref{app:pert-formula}]{Supplementary}
\begin{equation}
\label{PerturbationObservables}
    \kappa_n\left[\Tr \mathcal{O}\rho\right]= (-1)^n\frac{\partial^n\ln Z_K(M+\lambda\mathcal{O})}{\partial\lambda^n}\eval_{\lambda=0},
\end{equation}
where $\lambda$ is a small parameter. 

\begin{figure}[h]
    \centering
    \includegraphics[width=0.85\linewidth]{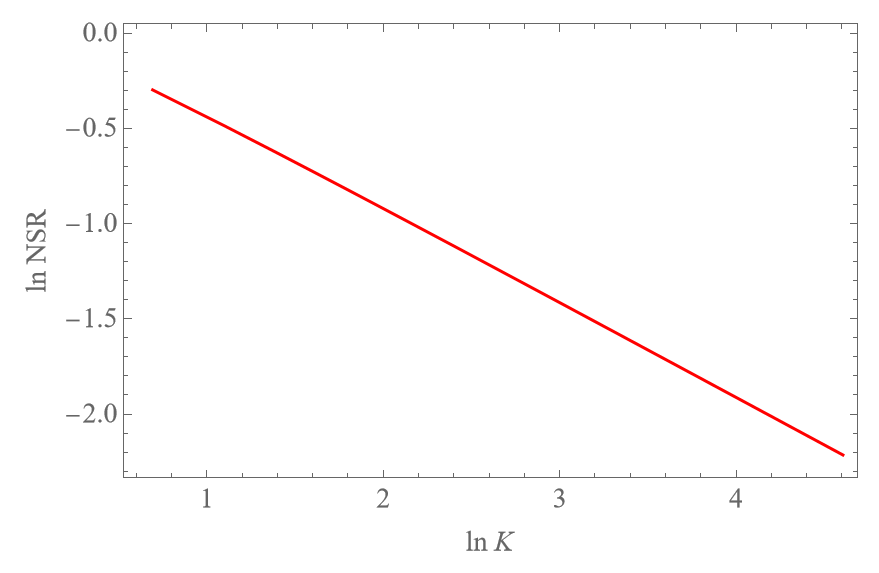}
    \caption{\small Log-log plot of the noise-to-signal ratio $\frac{\sqrt{\kappa_2}}{|\kappa_1|}$ of the energy $\Tr H\rho$ as a function of $K$ for a single-qubit subsystem, at $\beta = \Delta = 1$.}
    \label{NSRConc}
\end{figure}

Let $H = \frac{\Delta}{2}\sigma_z$. Taking the constraint matrix $M$ to be traceless for simplicity, we have $M = \frac{m}{2}\sigma_z$, with the parameter $m$ determined by Eq.~\eqref{ConstraintEqns}. As we show in~\cite[][Sec.~\ref{app:partition-function}]{Supplementary}, the partition function becomes
\begin{equation}
\label{QubitPartitionFunction}
    Z_K(m) = \frac{\sqrt{\pi}\,\Gamma(2K)}{\Gamma(K)} \,m^{\frac{1}{2}-K}\, I_{K-\frac{1}{2}}\left(\frac{m}{2}\right),
\end{equation}
where $I_\nu(\cdot)$ denotes the $\nu^{\text{th}}$ order modified Bessel function of the first kind. From Eq.~\eqref{PerturbationObservables}, the variance in the energy $\Tr H\rho$ is
\begin{equation}
\label{EnergyVarianceQubit}
    \kappa_2\left[\Tr H\rho \right] = \frac{\Delta^2}{4}\left[1-\frac{4K}{m}\frac{I_{K+\frac{1}{2}}}{I_{K-\frac{1}{2}}}-\left(\frac{I_{K+\frac{1}{2}}}{I_{K-\frac{1}{2}}}\right)^2\right],
\end{equation}
where the arguments $\frac{m}{2}$ of the Bessel functions have been omitted for brevity. In Fig.~\ref{NSRConc} we verify power-law ($K^{-1/2}$) concentration around the thermal average $\kappa_1$ in the limit of a large environment. 

\onecolumngrid
\newpage
\addtocontents{toc}{\protect\TocHideEnd}

\appendix

\setcounter{secnumdepth}{3}

\makeatletter
\@removefromreset{equation}{section}
\makeatother
\renewcommand{\theequation}{\arabic{equation}}
\setcounter{equation}{0}

\renewcommand{\thesection}{\Roman{section}}
\renewcommand{\thesubsection}{\thesection.\arabic{subsection}}
\renewcommand{\thesubsubsection}{\thesubsection.\arabic{subsubsection}}

\makeatletter
\renewcommand{\p@subsection}{}
\renewcommand{\p@subsubsection}{}
\makeatother

\begin{center}
    {\large\bfseries Supplemental material for: \\
    Maximum-Entropy Ensemble of Quantum Subsystems with a Fixed Average State}\\[0.5em]
    {\normalsize Charlie Shakeshaft, Christopher David White, Harry J. D. Miller}
\end{center}

\begin{bibunit}[apsrev4-2]

\setcounter{tocdepth}{2}
\tableofcontents

\vspace{1em}
\section{Maximum-entropy ensemble}
\label{AppA}
\subsection{Maximum-entropy measure}
\label{app:Max-Ent-measure}
We define probabilities here with respect to the Lebesgue measure $d\rho = \prod_{i=1}^d d\rho_{ii}\, \prod_{i<j}d\Re{\rho_{ij}}\,d\Im{\rho_{ij}}$ on $\mathcal{S}(\mathcal{H}_S)$. We solve for the probability distribution that maximises the relative entropy $-\int_{\mathcal{S}(\mathcal{H}_S)}P(\rho)\ln \left(\frac{P(\rho)}{P_K(\rho)}\right)\, d\rho$ with respect to the base distribution $P_K(\rho)$, subject to both the normalisation condition $\int_{\mathcal{S}(\mathcal{H}_S)} P(\rho)\, d\rho = 1$ and the first moment constraint $\int_{\mathcal{S}(\mathcal{H}_S)}\rho \, P(\rho) \, d\rho= \rho_0$. This leads to the functional
\begin{equation}
    \mathcal{L}\left[P(\rho)\right]= -\int_{\mathcal{S}(\mathcal{H}_S)} P(\rho)\ln\left(\frac{P(\rho)}{P_K(\rho)}\right)\, d\rho - \lambda_0\left(\int_{\mathcal{S}(\mathcal{H}_S)} P(\rho)\, d\rho -1\right)-\Tr \left[M\left(\int_{\mathcal{S}(\mathcal{H}_S)} \rho\, P(\rho)\, d\rho - \rho_0\right)\right].
\end{equation}
Setting the functional derivative $\delta \mathcal{L}/ \delta P = 0$, we obtain
\begin{equation}
    \ln\left(\frac{P(\rho)}{P_K(\rho)}\right) + 1 + \lambda_0 + \Tr(M\rho) = 0,
\end{equation}
which is solved by the normalised probability density
\begin{equation}
\label{MaxEntDensitySM}
    P_{\text{ME}}(\rho)= \frac{1}{Z_K(M)}\, e^{-\Tr M\rho}\, P_K(\rho).
\end{equation}
We note that the relative entropy over the flat measure $d\rho$ is mathematically equivalent to the \textit{ensemble entropy}, i.e. the Shannon entropy integrated over the measure $d\mu_K(\rho)$, and that the above solution is also obtained by maximising the latter quantity. 
\subsection{Partition function}
\label{app:partition-function}
We obtain the partition function $Z_K(M)$ by integrating the Gibbs-weighted measure $e^{-\Tr M\rho}d\mu_K(\rho)$ over the independent eigenvalues and eigenvectors of $\rho$. We follow the approach of Appendix C of~\cite{StatMechFixedEnergy} here, first evaluating the Haar integral using the \textit{Harish-Chandra-Itzykson-Zuber} (HCIZ) formula~\cite{HCIZMcSwiggen}
\begin{align}
    Z_K(M)&= \int d\mu_K(\vec{r})\times d\mu_{\text{Haar}}(U) \, e^{-\Tr M U D U^{\dagger}}\nonumber\\
    &= \int d\mu_K(\vec{r})\, \frac{\det\left[e^{-r_\nu m_\mu}\right]_{\nu, \mu}}{\Delta(D)\Delta(-M)},
\end{align}
where $\Delta(A) = \prod_{i<j}(a_i-a_j)$ is the \textit{Vandermonde determinant} of the matrix $A$ with spectrum $\{a_i\}_{i=1, \dots , d}$~\cite{MehtaRandomMatrices}. We then write the remaining eigenvalue integral as an inverse Laplace transform, giving
\begin{equation}
    Z_K(M)=\frac{1}{\Delta(-M)}\lim_{t\rightarrow 1}\mathcal{L}^{-1}_s\left[\Tilde{Z}_K(M,s)\right](t),
\end{equation}
where 
\begin{align}
    \Tilde{Z}_K(M,s)&=\frac{1}{d!}\int_0^\infty \mathcal{D}\rho \ \det\left[e^{-r_\nu(m_\mu +s)}\right]_{\nu, \mu}\det\left[r_\nu^\mu\right]_{\nu, \mu}\prod_{\nu=0}^{d-1}r_\nu^{K-d}\nonumber\\
    &=\frac{1}{d!}\int_0^\infty \mathcal{D}\rho \ \det\left[e^{-r_\nu(m_\mu +s)}\right]_{\nu, \mu}\det\left[r_\nu^{\mu+K-d}\right]_{\nu, \mu}.
\end{align}
We use the shorthand $\mathcal{D}\rho = dr_0 \cdots dr_{d-1}$ here. Using \textit{Andreief's integral formula}~\cite{ForresterAndreief} with $f_\mu(x)= e^{-x(m_\mu +s)}$ and $g_\mu(x)= x^{\mu + K -d}$, we obtain
\begin{align}
    \Tilde{Z}_K(M,s)&=\det\left[\int_0^\infty dx \ e^{-x(m_\mu + s)}x^{\nu+K-d}\right]_{\nu, \mu}\nonumber\\
    &= \det\left[\Gamma(1+\nu+K-d) \ (m_\mu+s)^{-(1+\nu+K-d)}\right]_{\nu, \mu}\nonumber\\
    &= \left(\prod_{\nu=0}^{d-1}\Gamma(1+\nu+K-d)\right)\det\left[ (m_\mu+s)^{-(1+\nu+K-d)}\right]_{\nu,\mu}\nonumber\\
    &= \left(\prod_{\nu=0}^{d-1}\Gamma(1+\nu+K-d)\right)\left(\prod_{\mu=0}^{d-1}(m_\mu+s)^{-K}\right)\det\left[ (m_\mu+s)^{-(1+\nu-d)}\right]_{\nu,\mu}.
\end{align}
We identify the rightmost determinant as the Vandermonde determinant $\Delta(M)$. Ignoring the leftmost factorial term (which is a constant), we obtain
\begin{equation}
    \Tilde{Z}_K(M,s)\propto \Delta(M)\frac{1}{\prod_{\mu=0}^{d-1}(m_\mu+s)^K},
\end{equation}
meaning the overall partition function is given by
\begin{equation}
\label{PartitionFunctionLaplaceIntegral}
    Z_K(M)=\Gamma(dK)\, \lim_{t\rightarrow 1}\mathcal{L}^{-1}_s\left[\prod_{k=0}^{d-1}\frac{1}{(m_k+s)^{K}}\right](t) \propto \frac{1}{2\pi i }\int_{\gamma -i\infty}^{\gamma+i\infty} ds \ e^s \prod_{k=0}^{d-1}\frac{1}{(m_k+s)^K},
\end{equation}
where the prefactor $\Gamma(dK)$ normalises according to the condition $Z_K(0)=1$. Now say the matrix $M$ has $L\leq d$ distinct eigenvalues, relabelled $m_1,\dots,m_L$ so that $m_k \neq m_{k'}$ $\forall k \neq k'$. Let the $k^{\text{th}}$ eigenvalue $m_k$ have multiplicity $n_k$, such that $\sum_{k=1}^L n_k = d$. The integral~\eqref{PartitionFunctionLaplaceIntegral} has $L$ poles in this case, where the $k^{\text{th}}$ pole is of order $n_kK$. Cauchy's residue theorem then gives
\begin{align}
    Z_K(M)&\propto \sum_k \text{Res}\left[\frac{e^z}{\prod_{j=1}^{L} (m_j +z)^{n_j K}}\ ; -m_k\right]\nonumber\\
    &= \sum_k \frac{1}{(n_k K - 1)!}\frac{d^{n_kK-1}}{dz^{n_kK-1}}\left[\frac{e^z}{\prod_{j=1, \ j\neq k}(m_j+z)^{n_jK}}\right]_{z=-m_k},
\end{align}
where the sum over $k$ is taken with respect to the distinct eigenvalues of $M$. The $(n_k K-1)^{\text{th}}$ derivative of the expression in the square brackets is
\begin{align}
    &\frac{d^{n_kK-1}}{dz^{n_kK-1}}\left[\frac{e^z}{\prod_{j=1, \ j\neq k}(m_j+z)^{n_jK}}\right]_{z=-m_k}\nonumber\\&= e^{-m_k}\sum_{l=0}^{n_kK-1}{{n_kK-1}\choose l}\left[\sum_{\abs{\vec{l}}=l}l! \prod_{j\neq k}{{n_jK+l_j-1}\choose{l_j}}\frac{(-1)^{l_j}}{(m_j-m_k)^{n_jK+l_j}}\right],
\end{align}
where the shorthand $|\vec{l}|$ represents the sum $\sum_{j\neq k}l_j$ of $L-1$ non-negative integers $l_j$, and the inner sum runs over all such tuples with $|\vec{l}|=l$ (each $l_j$ is then automatically at most $n_kK-1$). Substituting this in and commuting terms through the sum, we arrive at the analytical expression
\begin{align}
    Z_K(M) &\propto \sum_k \sum_{l=0}^{n_kK-1}\frac{(-1)^l e^{-m_k}}{(n_kK-l-1)!}\sum_{\abs{\vec{l}}=l}\prod_{j\neq k}{{n_jK+l_j-1}\choose{l_j}}\frac{1}{(m_j-m_k)^{n_jK+l_j}}.
\end{align}
When $d=2$ and we fix $M$ to be traceless with gap $m$, the integrand in~\eqref{PartitionFunctionLaplaceIntegral} simplifies to a product of 2 poles of order $K$, resulting in
\begin{equation}
    Z_K(m) = \Gamma(2K) \, \lim_{t\rightarrow 1}\mathcal{L}_s^{-1}\left[\frac{1}{\left(s+\frac{m}{2}\right)^K\left(s-\frac{m}{2}\right)^K}\right](t) = \frac{\Gamma(2K)}{\Gamma(K)}\,\sqrt{\pi}\, m^{\frac{1}{2}-K}\,I_{K-\frac{1}{2}}\left(\frac{m}{2}\right).
\end{equation}
\subsection{Perturbative formula for $n^\text{th}$ cumulant of arbitrary observables}
\label{app:pert-formula}
Consider the perturbation $M \to M+\lambda \mathcal{O}$, with $\mathcal{O} = \mathcal{O}^\dagger$ the observable of interest and $\lambda$ a small parameter. The partition function transforms as
\begin{align}
    Z_K(M+\lambda\mathcal{O}) &= \int_{\mathcal{S}(\mathcal{H}_S)}d\mu_K(\rho) \ e^{-\Tr M\rho} \, e^{-\lambda\Tr \mathcal{O}\rho}\nonumber\\
    &= Z_K(M)\left[\frac{1}{Z_K(M)}\int_{\mathcal{S}(\mathcal{H}_S)}d\mu_K(\rho) \ e^{-\Tr M\rho} \, e^{-\lambda\Tr \mathcal{O}\rho}\right]\nonumber\\
    &= Z_K(M) \langle e^{-\lambda \Tr \mathcal{O}\rho}\rangle.
\end{align}
Taking the logarithm of both sides yields 
\begin{align}
    \ln Z_K(M+\lambda\mathcal{O}) &= \ln Z_K(M) + \ln \langle e^{-\lambda \Tr \mathcal{O}\rho}\rangle \nonumber\\
    &= \ln Z_K(M)+ K_{\mathcal{O}}(-\lambda),
\end{align}
where $K_\mathcal{O}(t) = \ln \langle e^{t \Tr \mathcal{O}\rho}\rangle$ is the cumulant generating function of the observable. The $n^{\text{th}}$ cumulant is extracted by
\begin{equation}
    \kappa_n\left[\Tr \mathcal{O}\rho\right]= \frac{\partial^n}{\partial t^n}K_{\mathcal{O}}(t)\eval_{t=0}.
\end{equation}
It follows that 
\begin{equation}
    \frac{\partial^n}{\partial\lambda^n}\ln Z_K(M+\lambda\mathcal{O}) = (-1)^n K_\mathcal{O}^{(n)}(-\lambda)
\end{equation}
which, when evaluated at $\lambda = 0$, becomes
\begin{equation}
    \frac{\partial^n}{\partial\lambda^n}\ln Z_K(M+\lambda\mathcal{O})\eval_{\lambda=0} = (-1)^n \kappa_n\left[\Tr \mathcal{O}\rho\right].
\end{equation}
\section{Ensemble equivalence in asymptotic limit}
\label{app:secIII}
\subsection{Probability density of the FTCWE}
\label{app:FTCWE-density}
We derive the density of the FTCWE from the CWE density~\eqref{CWEDensity} by writing $W = t\rho$, where $t = \Tr W$, and then integrating out the trace $t$. The Jacobian for the transformation from $P(W)$ to the joint distribution $P(t,\rho)$ is  $dW = t^{d^2-1}dtd\rho$, and the determinant term factorises as 
\begin{equation}
    \det (W)^{K-d} = t^{d(K-d)}\det(\rho)^{K-d}.
\end{equation}
The resulting distribution is
\begin{equation}
    P(t,\rho)
    =\frac{K^{dK}}{\Gamma_d(K)\det(\Sigma)^K}\,
    t^{dK-1}
    \det(\rho)^{K-d}
    e^{- K t\Tr(\Sigma^{-1}\rho)},
\end{equation}
which we marginalise over $t$ to obtain
\begin{align}
    P(\rho)
    &=\frac{K^{dK}}{\Gamma_d(K)\det(\Sigma)^K}\,
    \det(\rho)^{K-d}
    \int_0^\infty
    t^{dK-1}
    e^{-Kt\Tr(\Sigma^{-1}\rho)}
    d t \nonumber\\
    &= \frac{\Gamma(dK)}
    {\Gamma_d(K)\det(\Sigma)^K}
    \frac{\det(\rho)^{K-d}}
    {\left[\Tr(\Sigma^{-1}\rho)\right]^{dK}},
\end{align}
where we have used the integral identity
\begin{equation}
    \int_0^\infty t^{dK-1}e^{-at}\,d t
    =
    \frac{\Gamma(dK)}{a^{dK}}
\end{equation}
with $a = K\Tr\Sigma^{-1}\rho$.
\subsection{Maximum-entropy, FTCWE, and POVM ensembles}
\label{app:Max-Ent-to-FTCWE}
In the following, we assume without loss of generality that the reference state $\rho_0$ is full-rank; if it is not, we can restrict our attention to the subspace of eigenvectors of $\rho_0$ with nontrivial eigenvalue
and take $d = \mathrm{rank}\; \rho_0$.

We consider four random-matrix ensembles in $\C^{d\times K}$:
maximum-entropy density matrices (ME),
\textit{correlated Wishart ensemble} (CWE) matrices,
\textit{fixed-trace correlated Wishart ensemble} (FTCWE) density matrices,
and density matrices resulting from the POVM protocol in the main text.
Each of these is parametrised by a positive-definite matrix in $\C^{d\times d}$%
---for the ME ensemble, a Lagrange multiplier $M \in \mathds C^{d \times d}$;
for the CWE and FTCWE, a correlation matrix $\Sigma$;
and for the POVM ensemble, a reference state $\rho_0$.
In this section  we take
\begin{align}
\label{SigmaGauge}
\Sigma = KM^{-1} = \rho_0\;.
\end{align}
Note that this requires appropriately choosing the gauge of $M$.
Note also that this identification is only strictly appropriate in the large-$K$ limit:
the CWE has $\Sigma$ as its exact average, but the FTCWE, ME, and ($\rho_0$-centered) POVM averages only approach $\Sigma$ in the large-$K$ limit.
In our more rigorous Sec.~\ref{sss:integrability} we relax this assumption:
we do not identify $M$ and $\Sigma$.
In general we take
\begin{align}
    \Tr\Sigma = 1\;;
\end{align}
when $\Sigma = KM^{-1}$ this can be arranged by appropriately choosing the gauge of $M$.

Three of the four ensembles---the ME, FTCWE, and POVM ensembles---can be understood as the images of three ensembles on the complex sphere $S^{2dK - 1} = \{ \omega \in \mathds C^{d\times K}\ :\ \|\omega\|_2 = 1\}$ under the same map
\begin{align}
\rho = \omega \omega^\dagger\;.
\end{align}

In this subsection we show that the ME, the FTCWE, and the POVM ensembles are equivalent.
First, in Sec.~\ref{sss:pdfs}, we relate the probability density functions (PDFs) of the ME, FTCWE, and POVM ensembles on the sphere variables $\omega$.
We show that each PDF depends only on
\begin{equation}
\label{qomega}
q(\omega) = \frac 1 {dK} \Tr[M\omega\omega^\dagger]
\end{equation}
---in particular, that in each case the pdf is\footnote{
Here and elsewhere, we implicitly use the Haar measure on the sphere as our base measure---our PDFs are Radon-Nikodym derivatives with respect to that measure.
}
\begin{align}
p(\omega) \propto e^{-dK I(q(\omega))}
\end{align}
for some rate function $I(q)$ with a minimum near $q = 1$.
This informally establishes that the three ensembles are (in the large-$K$ limit) equivalent to the ``microcanonical'' ensemble with $q = 1$, described by the PDF
\begin{align}
p(\omega) = \delta\big(q(\omega) - 1\big)\;.
\end{align}
This is analogous to the equivalence of classical microcanonical and canonical ensembles.

Then in Sec.~\ref{sss:integrability} we rigorously establish a stronger result: that the ME, FTCWE, and POVM ensembles are (again in the large-$K$ limit) equivalent to the much more tightly constrained ``generalized microcanonical'' ensemble.
To understand the generalized microcanonical ensemble, decompose
\begin{align}
q(\omega) = \frac 1 d \sum_{j = 1}^d q_j(\omega),\quad q_j(\omega) = \frac {m_j} K \sum_{k = 1}^K |\omega_{jk}|^2\ .
\end{align}
The $q_j$ play the role of the extensive set of conserved quantities in an integrable Hamiltonian, 
and the generalized microcanonical ensemble is similar to the generalized Gibbs ensemble of such a Hamiltonian.
In the generalized microcanonical ensemble, the ``conserved quantities'' $q_j$ are fixed:
\begin{align}
p(\omega) = \prod_{j = 1}^{d} \delta\big(q_j(\omega) - 1)\;.
\end{align}
This is broadly analogous to the microcanonical version of the generalized Gibbs ensemble familiar from the study of integrable systems.
Concretely, we show that expectation values of Lipschitz functions in the FTCWE, ME, and POVM ensembles approach those of the generalized microcanonical ensemble as $K \to \infty$.

\subsubsection{Relationship of maximum-entropy, FTCWE, and POVM probability density functions}\label{sss:pdfs}

Maximum-entropy density matrices are $\rho_{\ME}$ distributed according to the PDF~\eqref{MaxEntDensitySM}, with the Lagrange multiplier $M$ chosen so the ensemble average matches the reference state $ \langle \rho\rangle_{\ME} = \rho_0$. The ensemble is Gibbs-weighted with respect to the base distribution $P_K(\rho)$ induced by the partial trace on Haar states of dimension $dK$, so our $\rho_\ME$ come from a distribution over the sphere:
\begin{align}
    \rho_{\ME} = \Omega \Omega^\dagger\;,
\end{align}
where $\Omega$ is a random variable on the sphere  $S^{2dK-1} =\{\omega \in \C^{d\times K} \; :\; \|\omega\|_2 = 1\}$.
In this case $\Omega$ is drawn from the complex Bingham distribution---that is,
the unit sphere with probability distribution 
\begin{align}\begin{split}
    p_{\ME}(\omega) &\propto e^{-\Tr[M \omega\omega^\dagger ]} \\
    &= e^{-dKq(\omega)}\;,
\end{split}\end{align}
with $q(\omega)$ defined as in Eq.~\eqref{qomega}.

Correlated Wishart matrices with correlation $\Sigma$ are
\begin{equation}
\label{CWEform}
    W_{\CWE} = \frac{1}{K}XX^\dagger\;, X \sim \CN_{d\times K}(0,\Sigma, I_K),
\end{equation}
where $\CN_{d\times K}(0,\Sigma, I_K)$ is the distribution over $\C^{d\times K}$ with probability density function
\begin{equation}
    p_{\CWE}(X) \propto e^{-\Tr[\Sigma^{-1} XX^\dagger]}.
\end{equation}
Note that this $X$ takes values in all of $\C^{d\times K}$,
while the variable $\Omega$ of the maximum-entropy distribution takes values on the unit sphere.
The CWE is (induced by) a Gaussian distribution over the $X$, and one can check that
\begin{align}
    \begin{split}
    \langle X\rangle &= 0\\
    \langle W_\CWE\rangle &= \frac{1}{K}\langle XX^\dagger\rangle\\
    &= \Sigma \;.
    \end{split}
\end{align}
Note also that $W_{\CWE}$ is positive semidefinite but not a density matrix:
the trace varies around $\Tr\Sigma$.

Fixed-trace correlated Wishart matrices with correlation $\Sigma$ are now correlated Wishart matrices with the trace set to 1:
\begin{equation}
    \rho_{\FTCWE} = \frac{W_{\CWE}}{\Tr W_{\CWE}}.
\end{equation}
We can put FTCWE density matrices on an equal footing with the maximum-entropy density matrices by rearranging
to write them in terms of a random variable $\Phi$ on the sphere:
\begin{align}
\label{FTCWEsphere}
    \begin{split}
        \rho_{\FTCWE} &= \Phi \Phi^\dagger\;,\\
        \Phi &= X / \|X\|_2 \in S^{2dK-1}\;,\\
        X &\sim \CN_{d\times K}(0, \Sigma, I_K)\; .
    \end{split}
\end{align}
We can write an explicit probability density function for the $\Phi$
because $\Phi$ is precisely the unit-sphere component of a polar decomposition of the CWE variable $X = R\Phi, R \in [0,\infty)$:
\begin{align}
\begin{split}
    p(\varphi) &= \int dr\; r^{2dK-1} e^{-r^2 \Tr[\Sigma^{-1}\varphi\varphi^{\dagger}]} \\
    &\propto \left(\Tr[\Sigma^{-1}\varphi\varphi^{\dagger}]\right)^{-dK}\;.
    \end{split}
\end{align}
When $M \succ 0$ we can take $\Sigma = KM^{-1}$, and this becomes
\begin{align}
p_\FTCWE(\varphi) \propto q(\varphi)^{-dK}\;.
\end{align}

The FTCWE and the maximum-entropy ensemble, then, come from two distributions on the unit sphere,
with probability density functions that depend only on $q$.
\begin{align}
\begin{split}
p_\ME(\varphi) &\propto e^{-dKq(\varphi)}\\
p_\FTCWE(\varphi) &\propto q(\varphi)^{-dK}\;.
\end{split}
\end{align}
It remains only to show that these are both close to the microcanonical ensemble
\begin{align}
    p_{\mu c}(\varphi) &\propto \delta(q(\varphi) - 1)\;.
\end{align}
To see that this is the case for the FTCWE, rewrite $q$ in terms of a random variable $\Omega$ isotropically distributed on the sphere.
Above we wrote FTCWE matrices in terms of a random variable $\Phi$ that was the projection of an anisotropic Gaussian onto the sphere (Eq.~\eqref{FTCWEsphere}). 
But the CWE $X$ can be written 
\begin{align}
X = \Sigma^{1/2} Y,\quad Y \sim \mathcal{CN}_{d\times K}(0, I_d, I_K)
\end{align}
(that is, the elements $Y_{ij}$ are i.i.d. $\sim \mathcal{CN}(0,1)$). Then the FTCWE sphere variable is
\begin{align}\label{eq:phi-gaussian-Y}
\Phi = \frac{\Sigma^{1/2} Y} { \sqrt{\Tr[\Sigma YY^\dagger]}}\; , 
\end{align}
and 
\begin{align}
\begin{split}
q(\Phi) &= \frac 1 {d} \frac {\Tr[YY^\dagger]}{\Tr[\Sigma YY^\dagger]}\\
&= \frac 1 {d} \frac 1 {\Tr[\Sigma \Omega \Omega^\dagger]}
\end{split}
\end{align}
where $\Omega = Y / \|Y\|_2$ is isotropically distributed on the unit sphere.
Consider $\C^{d\times K} \equiv \C^{dK} \equiv \R^{2dK}$ as a vector space and expand it in an eigenbasis of $\Sigma \otimes I_{K} \in \mathds C^{dK \times dK}$.
Considered as an operator on $\R^{2dK}$, $\Sigma \otimes I_{K}$ has eigenvalues $t_i$; they consist of the eigenvalues of $\Sigma$ each with multiplicity $2K$.
Then
\begin{align}\label{eq:q-phi-dirichlet}
q(\Phi)^{-1} = d \sum_{i = 1}^{2dK} t_i \Omega_i^2.
\end{align}
The $\Omega_i^2$ come from the Dirichlet distribution of order $2dK$ with parameters $\alpha_i = 1/2$, so standard results on that distribution, together with short calculations, give that the mean of $q^{-1}$ is
\begin{align}
\begin{split}
\langle q(\Phi)^{-1}\rangle  &= \Tr\Sigma = 1
\end{split}
\end{align}
where we have used that $\Tr\Sigma = 1$ by assumption,
and the variance of $q^{-1}$ is
\begin{align}\label{eq:FTCWE-q-var}
\mathrm{var}[q(\Phi)^{-1}] &= \frac d {dK + 1} \left(\Tr[\Sigma^2] - \frac 1 d\right)\; . 
\end{align}
Since $\Tr\Sigma = 1$ and $\Sigma$ is positive-definite, $\Tr[\Sigma^2] \le 1$ and 
\begin{align}
\mathrm{var}[q(\Phi)^{-1}] &= \mathcal{O}(K^{-1}) \ll \langle q(\Phi)^{-1}\rangle^2
\end{align}
is small, so the distributions of both $q(\Phi)^{-1}$ and $q(\Phi)$ are peaked at $q(\Phi) = 1$ with width given by \eqref{eq:FTCWE-q-var};
that width is $\mathcal{O}(K^{-1/2})$.
Consequently
\begin{align}
    p_{\FTCWE}(\varphi) \approx \delta(q(\varphi) - 1)
\end{align}
in the limit of large $K$.

To see that the ME ensemble also approaches the microcanonical ensemble,
note that the ratio
\begin{align}\label{eq:me-ftcwe}
\frac{p_{\ME}(\varphi)}{p_\FTCWE(\varphi)} \propto \exp\left(-dK\left[q(\varphi) - \ln q(\varphi)\right]\right)
\end{align}
is also sharply peaked at $q = 1$ with width $\propto (dK)^{-1/2}$, so likewise
\begin{align}
    p_{\ME}(\varphi) \approx \delta(q(\varphi) - 1)
\end{align}
in the limit of large $K$.

We have not constructed a simple argument that the POVM ensemble is peaked in $q$, but this follows from convergence to the generalized microcanonical ensemble (Prop.~\ref{prop:povm-ftcwe}).

\subsection{ME, FTCWE, and POVM ensembles converge to the generalized microcanonical ensemble}\label{sss:integrability}

In this section we prove that the ME, FTCWE, and POVM ensembles all converge to the generalized microcanonical ensemble.

We first consider the ME and FTCWE.
We show that if the weights of the ensembles on the eigenspaces of the covariance matrix $\Sigma$ approach the eigenvalues of $\Sigma$, then the ensembles are close in Wasserstein distance to the generalized microcanonical ensemble, so expectation values of Lipschitz functions converge (Lemma \ref{lem:wass-eigenspace}).
We then check that the ME and FTCWE ensembles have the relevant properties (Props. \ref{prop:ftcwe-amplitude} and \ref{prop:me-amplitude}).
Heuristically, this comes about via central limiting behavior:
in each distribution, the weight on each eigenspace of $\Sigma$ is the sum of $K$ random variables,
so the standard deviation is small like $1/\sqrt K$.
Because the $K$ random variables are not independent, however, some care must be taken.
With the relevant properties established, convergence follows immediately (Prop.~\ref{prop:me-ftcwe-gmu}).

We then turn to the POVM.
We show directly that it is close to the FTCWE in total variation distance, hence Wasserstein distance and expectation of Lipschitz functions (Prop.~\ref{prop:povm-ftcwe}); 
convergence to the generalized microcanonical ensemble comes as a corollary.

\begin{defn}
    Take a covariance matrix $\Sigma \in \C^{d\times d}$, $0 \prec \Sigma$, $\Tr \Sigma = 1$; write its eigenvalues $s_j$.

    The \textit{generalized microcanonical ensemble} on the sphere $S^{2dK-1}$ with covariance matrix $\Sigma$ is the distribution with pdf (in a $\Sigma$ eigenbasis)
    \begin{align}
    p(\omega) = \prod_{j = 1}^{d-1} \delta\left(s_j - \sum_k |\omega_{jk}|^2 \right)
    \end{align}
    ---that is, the $\Sigma$ eigenspace weights are all fixed to precisely the $\Sigma$ eigenvalues.
    \footnote{There are only $d-1$ because those delta functions, together with $\sum_j s_j = \sum_j \sum_k |\omega_{jk}|^2 = 1$, fix the last $s_d = \sum_{k} |\omega_{j = d,k}|^2$.}
\end{defn}

\begin{rem}
Note that this definition is slightly different from that of the previous section: they are only identical in the large-$K$ limit, where $\Sigma \approx KM^{-1}$.
\end{rem}

\begin{lemma}\label{lem:wass-eigenspace}
    Consider a random variable $\Omega \in S^{2dK-1} \subset \mathds C^{d\times K}$ such that each row of $\Omega$ is separately invariant in distribution under right unitary multiplication.
    Take (as elsewhere in this section) a Hermitian matrix $0 \prec \Sigma \in \mathds C^{d\times d}$, $\Tr \Sigma = 1$, with eigenvalues and eigenvectors $s_i, \ket i$.
    Suppose that the amplitudes of $\Omega$ in those eigenspaces
    \begin{align}
    \sqrt{R_i} \equiv \sqrt{\bra{i} \Omega \Omega^\dagger \ket{i} }
    \end{align}
    are close to $\sqrt{s_i}$ in the sense that
    \begin{align}
    \E\left[\sum_i \left(\sqrt{R_i} - \sqrt{s_i}\right)^2\right] \le  B/K
    \end{align}
    for some $B$.
    Then $\Omega$ is close to the generalized microcanonical ensemble in the following two senses:
    \begin{enumerate}
        \item The Wasserstein distance (of the underlying measures) is 
        \begin{align}
        W_1 \le \sqrt{B/K}
        \end{align}
        (where the underlying metric is the Frobenius, i.e. Euclidean, metric on $\C^{d\times K}$).
        \item Consequently, expectation values of Lipschitz functions agree: if a function 
        \begin{align}
            f : \C^{d \times d} \to \mathds R
        \end{align}
        is Lipschitz with coefficient $L$, then
        \begin{align}
        \Big| \E[f(\Omega\Omega^\dagger)] - \E[f(\Omega_{g\mu}\Omega_{g\mu}^\dagger)]\Big| \le 2L \sqrt{B/K}
        \end{align}
        where we write $\Omega_{g\mu}$ for a random variable in the generalized microcanonical ensemble.
    \end{enumerate}
\end{lemma}

\begin{proof}

First consider the Wasserstein distance; expectation values of Lipschitz functions will follow via Kantorovich-Rubinstein duality.

Since $\Omega$ is $\C^{d\times K} \equiv \C^d \otimes \C^K$-valued
and invariant in distribution under rotations $U(K)$ on each eigenspace of $\Sigma$,
we can decompose it 
\begin{align}
\Omega = \sum_{j = 1}^d \sqrt{R_j} \ket{j} \otimes \ket{\Theta_j}
\end{align}
where $\ket{j} \in \C^d$ are eigenvectors of $\Sigma$, $\ket{\Theta_j}$ are independent $\C^K$-valued random variables from the Haar distribution on the sphere $S^{2K-1} = \{\ket{\theta} \in \C^K : \braket{\theta}{\theta} = 1\}$, and $\sqrt{R_j}$ are positive real amplitudes, i.e. $\R$-valued random variables.

The map
\begin{align}
\sum_{j = 1}^d \sqrt{r_j} \ket{j}\otimes\ket{\Theta_j}\quad \mapsto \quad \sum_j \sqrt{s_j} \ket{j} \otimes \ket{\Theta_j }
\end{align}
gives a coupling between the distribution of $\Omega$ and the generalized microcanonical distribution.
The Wasserstein distance is then
\begin{align}\label{eq:W1-bound-roots-cpts}
\begin{split}
W_1^2 &\le
\mathds E
\left[ \left\|
\sum_{j = 1}^d \sqrt{R_j} \mathbf e_j \otimes \Theta_j\quad
- \quad \sum_j \sqrt{s_j} \mathbf e_j \otimes \Theta_j 
\right\|_2^2 \right] \\
&= \E\left[\sum_j \left(\sqrt{R_j} - \sqrt{s_j}\right)^2\right]\\
&\le B/K\;.
\end{split}
\end{align}

Now take a function 
\begin{align}
    f : \C^{d \times d} \to \mathds R
\end{align}
Lipschitz with coefficient $L$.
The map on the sphere
\begin{align}
    \begin{split}
        g &: S^{2dK-1} \subset \C^{d\times K} \to \mathds R \\
        g &:\omega \mapsto f(\omega\omega^\dagger)
    \end{split}
\end{align}
is likewise Lipschitz with coefficient $2L$,
precisely because of the restriction to the sphere---we can do this because $\Omega, \Omega_{g\mu}$ are sphere-valued.
The desired result follows immediately from Kantorovich-Rubinstein duality.

\end{proof}
\begin{rem}
    It is clear that the CWE with correlation matrix $\Sigma$ satisfies the hypotheses of the lemma:
    each squared amplitude is the sum of i.i.d. Gaussian random variables, so it is close to $\sqrt{s_i}$ in the appropriate sense.
    But in the FTCWE and ME ensembles, the terms in the amplitude sum are not independent.
    The bulk of this subsection is given over to technical tools to route around that lack of independence.
\end{rem}

\begin{prop}\label{prop:ftcwe-amplitude}
The FTCWE has eigenspace amplitudes close to $\sqrt{s_i}$: 
if $\Omega \sim \FTCWE(\Sigma, K)$ with $0 \prec \Sigma$ and $\Tr \Sigma = 1$,
and we write
\begin{align}
\sqrt{R_i} \equiv \sqrt{\bra{i} \Omega \Omega^\dagger \ket{i} }\;,
\end{align}
then
\begin{align}
    \E\left[\sum_i (\sqrt{R_i} - \sqrt{s_i})^2\right] \le 4/K\;.
\end{align}
\end{prop}
\begin{proof}
The proof will end up resting on the central limit theorem for the CWE from which the FTCWE is constructed,
but we must deal with the projection onto the sphere.

Since the FTCWE variable $\Omega$ is sphere-valued and $\Tr \Sigma = 1$,
we can write unit vectors
\begin{align}\label{eq:ftcwe-unit-vecs}
\begin{split}
a,b &\in \mathds R^d \\
a_i &= \sqrt{R_i}\\
b_i &= \sqrt{s_i}\;;
\end{split}
\end{align}
the quantity of interest is then
\begin{align}
   \E\left[\sum_i (\sqrt{R_i} - \sqrt{s_i})^2\right] \equiv \E[\|a - b\|_2^2]\;.
\end{align}

To bound that expectation value
decompose the FTCWE variable $\Omega$ in terms of i.i.d. Gaussians, just as in the derivation of \eqref{eq:q-phi-dirichlet} above:
\begin{align}
|\Omega_{ij}|^2 = \frac {s_i |Y_{ij}|^2}{\sum_{ij} s_i |Y_{ij}|^2}
\end{align}
where $Y_{ij} \sim \CN(0,1/K)$ i.i.d.%
\footnote{
Note that the variance here differs by a constant factor from that of the previous section.
Both here and there we are free to choose that constant factor:
constant factors are related by a rescaling.
}
The weights $R_i$, i.e. the components of $a$ in \eqref{eq:ftcwe-unit-vecs}, are then
\begin{align}
R_i = \frac {s_i \sum_j |Y_{ij}|^2}{\sum_{ij} s_i |Y_{ij}|^2}
\end{align}
so we can write the vector $a$ as
\begin{align}
a = \frac{a'}{\|a'\|_2}\;,\quad {a'}^2_i = s_i \sum_{j = 1}^K |Y_{ij}|^2\;.
\end{align}
Since $b$ is a unit vector,
\begin{align}
\Big\| a' / \|a'\|_2 - b \Big\|_2 \le 2 \|a' - b\|_2\;,
\end{align}
so
\begin{align}
\begin{split}
\E[\|a - b\|_2^2] &\le 4\E[\|a' - b\|_2^2] \\
&\le 4\E\left[ \sum_{i = 1}^d s_i \left( \sum_{j = 1}^K |Y_{ij}|^2 - 1\right)^2 \right]\\
&= 4/K
\end{split}
\end{align}
using that $(\sqrt x - 1)^2 \le (x - 1)^2$ for $x \ge 0$, expanding and taking Gaussian expectation values, and then applying $\sum_i s_i = \Tr \Sigma = 1$.
\end{proof}

\begin{prop}\label{prop:me-amplitude}
Consider the ME with matrix $M$ chosen so $\E \Omega_{\ME}\Omega_{\ME}^\dagger = \Sigma$. The ME has eigenspace amplitudes close to $\sqrt{s_i}$:
if $\Omega_\ME \sim \ME(\Sigma, K)$, $0 \prec \Sigma$ and $\Tr \Sigma = 1$, and we write
\begin{align}
\sqrt{R_i} \equiv \sqrt{\bra{i} \Omega_\ME \Omega_\ME^\dagger \ket{i} }\;,
\end{align}
then
\begin{align}
    \E\left[\sum_i (\sqrt{R_i} - \sqrt{s_i})^2\right] \le 1/K\;.
\end{align}
\end{prop}

\begin{rem}
    Note that this does not assume $\Sigma = KM^{-1}$---instead we choose a slightly different $M$ to match $\Sigma$.
\end{rem}

\begin{rem}
The Lagrange multiplier $M$ parametrizing the ME ensemble commutes with the reference state, in this case $\Sigma$.
To see this, note that the ME maximises the ensemble entropy subject to the constraint $\mathds E[\Omega_{\ME}\Omega_{\ME}^\dagger] = \Sigma$;
both the constraint and the optimization target (the entropy) are invariant under the unitary
\begin{align}
\Omega_{\ME} \to e^{i\theta \Sigma} \Omega_{\ME}\;.
\end{align}
\end{rem}

\begin{proof}
The PDF of the ME ensemble is
\begin{align}
p(\omega) \propto e^{-dKq(\omega)}\;.
\end{align}
Because $M$ commutes with $\Sigma$, we can decompose $\omega$ into amplitudes and angles
\begin{align}
\omega = \sum_j \sqrt{r_j} \ket j \otimes \ket{\Theta_j}\;,\quad \sum_j r_j = 1\;,
\end{align}
where the $\ket j$ are eigenvectors of $M$, hence also $\Sigma$.
The PDF is then
\begin{align}
p\left(\sum_j \sqrt{r_j} \ket j \otimes \ket{\Theta_j}\right) \propto \delta\left(\sum_j r_j - 1\right)\prod_j e^{-m_jr_j} 
\end{align}
independent of the angles $\ket{\Theta_j}$.
Integrating the angles out gives a $(\sqrt r)^{2K-1}$ factor for each $j$, so the marginal distribution on the amplitudes is 
\begin{align}
p(\sqrt{r}) = \delta\left(\sum_j r_j - 1\right)\prod_j r_j^{K-1/2} e^{-m_j r_j}\;.
\end{align}
Note that this is the joint pdf of the amplitudes $\sqrt{r_j}$, not the weights $r_j$;
the joint pdf of the weights is 
\begin{align}
p(r) = \delta\left(\sum_j r_j - 1\right) \prod_j r_{j}^{K-1} e^{-m_j r_j}\;.
\end{align}

We seek to show that the weights $R_j$ are close to the $s_j$.
By construction $\E R_j = s_j$, so we can accomplish this by showing that $\mathrm{var}\;R_j$ is small,
then manipulating to get precisely the right form.
Were the $R_j$ independent the small variance would be a straightforward application of the central limit theorem, but they are not independent---the delta function, i.e. the constraint that the ME sphere variables in fact be on the sphere, breaks the independence.

Instead, to show that the variance is small, use some facts about log-concave functions.
The pdf of the marginal distribution of a single $R_j$ is 
\begin{align}
p(r_j) = r_j^{K-1} h_j(r_j)\;,
\quad h_j(r_j) =  e^{-m_j r_j} \int \prod_{k \ne j} \left[dr_k r_{k}^{K-1} e^{-m_k r_k}\right] \delta\left(\sum_k r_k - 1\right)\;.
\end{align}
The integral is the iterated convolution of log-concave functions, so by a theorem of Pr\'ekopa \cite{prekopa1973} each $h_j$ is itself log-concave.
By a result of Bobkov and Madiman \cite[Sec.~4]{bobkovConcentrationInformationData2011}, then,
\begin{align}
\mathrm{var}\; R_j \le \frac{(\E R_j)^2}{K} = s_j^2/K
\end{align}
since $\mathds E R_j = s_j$.

To turn this into a bound on $\|\sqrt r - \sqrt s\|_2$, note that
\begin{align}
(\sqrt{r_j} - \sqrt{s_j})^2 = \frac{(r_j - s_j)^2}{(\sqrt{r_j} + \sqrt{s_j})^2} \le \frac{(r_j - s_j)^2}{s_j}
\end{align}
so the quantity we seek to bound is
\begin{align}
\begin{split}
 \E\left[\sum_j (\sqrt{R_j} - \sqrt{s_j})^2\right] &\le \sum_j \frac 1 {s_j} \mathds E[(R_j - s_j)^2]\\
 &\le \sum_j \frac{s_j} K\\
 &= 1/K\;,
\end{split}
\end{align}
recognizing $\E[(R_j - s_j)^2] = \mathrm{var}\; R_j \le s_j^2/K$.
\end{proof}

\begin{prop} \label{prop:me-ftcwe-gmu}
Write constants
\begin{align}
\begin{split}
B_{\FTCWE} &= 4\\
B_{\ME} &= 1 \;.
\end{split}
\end{align}
Then
\begin{enumerate}
    \item The Wasserstein distances between the generalized microcanonical ensemble and the FTCWE and ME ensembles are
    \begin{align}
    W_1 \le \sqrt{B_{\{\FTCWE,\ME\}}/K}
    \end{align}
    respectively 
    (once again the underlying metric is the Frobenius, i.e. Euclidean, metric on $\C^{d\times K}$).
    \item Consequently expectation values of Lipschitz functions agree: if a function 
    \begin{align}
        f : \C^{d \times d} \to \mathds R
    \end{align}
    is Lipschitz with coefficient $L$, then
    \begin{align}
    \Big| \E[f(\Omega\Omega^\dagger)] - \E[f(\Omega_{g\mu}\Omega_{g\mu}^\dagger)]\Big| \le 2L\sqrt{B_{\{\FTCWE,\ME\}}/K}
    \end{align}
    where $\Omega \sim \{\FTCWE, \ME\}$ and we write $\Omega_{g\mu}$ for a random variable in the generalized microcanonical ensemble.
\end{enumerate}
\end{prop}
\begin{proof}
Propositions \ref{prop:ftcwe-amplitude} and \ref{prop:me-amplitude} establish that the FTCWE and ME satisfy the hypotheses of the Lemma \ref{lem:wass-eigenspace}, which gives the desired result.
\end{proof}

\begin{rem}
Strictly speaking, we have shown that the FTCWE and the ME ensemble have small Wasserstein distance to the generalized microcanonical distribution as distributions over the sphere $S^{2dK-1}$.
Because the pushforward
\begin{align}
\omega \mapsto \omega\omega^\dagger
\end{align}
is Lipschitz with constant 2, however, this gives that they have small Wasserstein distance as distributions over density matrices in $\C^{d\times d}$.
\end{rem}

\begin{prop}\label{prop:povm-ftcwe}
Consider the POVM and FTCWE ensembles on the sphere $S^{2dK-1} \subset \C^{d\times K}$
with covariance matrix $0 \prec \Sigma$, $\Tr \Sigma = 1$.
The total variation distance between the two ensembles is
\begin{align}
d_{\mathrm{TV}} \le \sqrt{\frac 1 4 \frac d {dK+1} \left(\Tr[\Sigma^2] - \frac 1 d\right)} < \frac 1 {2\sqrt K}
\end{align}
so the Wasserstein distance is
\begin{align}
W_1 \le \sqrt{\frac d {dK+1} \left(\Tr[\Sigma^2] - \frac 1 d\right)} < 1/\sqrt{K}\;;
\end{align}
and Lipschitz functions on density matrices agree: if
\begin{align}
    f : \C^{d \times d} \to \mathds R
\end{align}
is Lipschitz with coefficient $L$, then
\begin{align}
\Big| \E[f(\Omega_\POVM\Omega_\POVM^\dagger)] - \E[f(\Omega_{\FTCWE}\Omega_{\FTCWE}^\dagger)]\Big|
\le 2L \sqrt{\frac d {dK+1} \left(\Tr[\Sigma^2] - \frac 1 d\right)} < 2L/\sqrt{K}
\end{align}
where $\Omega_{\POVM, \FTCWE}$ are sphere variables in the POVM and FTCWE ensembles.
\end{prop}

\begin{proof}
The ratio of PDFs of the two distributions is%
\footnote{
In writing the PDFs, we have so far ignored constant factors,
so one might think that we can only write the ratio up to constant factors.
But the normalization constraint on the ratio of PDFs
\begin{align}
1 = \E\left[ \frac{p_{\POVM}(\Omega_\FTCWE)}{p_{\FTCWE}(\Omega_\FTCWE)} \right]\;,
\end{align}
together with the fact that $\E[q(\Omega_\FTCWE)^{-1}] = 1$,
gives the constant factor.
}
\begin{align}
\frac{p_{\POVM}(\omega)}{p_{\FTCWE}(\omega)} = q(\omega)^{-1}\;,
\end{align}
so the total variation distance is
\begin{align}
\begin{split}
d_{\mathrm{TV}} &= \frac 1 2 \int d\omega\; p_{\FTCWE}(\omega) \left|\frac{p_{\POVM}(\omega)}{p_{\FTCWE}(\omega)} - 1\right| \\
&= \frac 1 2 \E\left[\left|q^{-1}(\Omega_\FTCWE) - 1 \right|\right]\;,
\end{split}
\end{align}
where $\Omega_\FTCWE$ is an FTCWE sphere variable.
Therefore
\begin{align}
\begin{split}
d_{\mathrm{TV}}^2
    &\le \frac 1 4 \E\left[\left|q^{-1}(\Omega_\FTCWE) - 1 \right|^2\right] \\
    &= \frac 1 4 \mathrm{var}\;q^{-1}(\Omega_\FTCWE) \\
    &\le \frac 1 4 \frac d {dK+1} \left(\Tr[\Sigma^2] - \frac 1 d\right) < \frac{1}{4K}
\end{split}
\end{align}
by the variance calculation of Sec.~\ref{sss:pdfs}.
Since the Hilbert-Schmidt diameter of the unit sphere is 2, the Wasserstein distance is
\begin{align}
W_1 \le 2d_{\mathrm{TV}} \le \sqrt{\frac d {dK+1} \left(\Tr[\Sigma^2] - \frac 1 d\right)} < 1/\sqrt K\;;
\end{align}
Lipschitz functions follow by the same Kantorovich-Rubinstein argument as in Lemma \ref{lem:wass-eigenspace}.
\end{proof}

\begin{cor}
    By the triangle inequality, the POVM ensemble is close like $O(1/\sqrt K)$ to the generalized microcanonical ensemble in Wasserstein distance and expectation values of Lipschitz observables (but not total variation distance).
\end{cor}

\subsection{Concentration of normalization}
\label{app:concentration-normalization}

\begin{prop}\label{prop:bernstein-trace} (Bernstein inequality for trace of CWE matrices)
    If $W \sim \CWE(\Sigma, K)$, $\Tr\Sigma = 1$, then
    \begin{align}
    P(|\Tr W - \Tr\Sigma| \ge t)  &\le 2 \exp\left[- \kappa \min\left(\frac{9Kt^2}{32\|\Sigma\|^2_2}, \frac {3Kt} {4 \|\Sigma\|_{\mathrm{op}} }\right)\right]\;.
    \end{align}
\end{prop}
\begin{cor}\label{cor:trace-concentration}
    If $\Sigma = \rho_0$, where $\rho_0$ is our reference state,
    \begin{align}
    P(|\Tr W - 1| \ge t)  &\le 2 \exp\left[- \kappa \min\left(\frac{9Kt^2}{32\Tr[\rho_0^2]}, \frac {3Kt} {4 \|\rho_0\|_{\mathrm{op}} }\right)\right]
    \end{align}
    i.e. the characteristic scale of fluctuations of $\Tr W$ is
    \begin{align}
    \begin{split}
     \sim K^{-1/2}e^{-S_2(\rho_0)/2}\quad\text{Gaussian core} \\
     \sim K^{-1}e^{-S_\infty(\rho_0)}\quad\text{exponential tail} \\
     \end{split}
    \end{align}
        where $S_2(\rho_0)$ and $S_\infty(\rho_0)$ are the second R\'enyi entropy and the min-entropy, respectively.
\end{cor}

\begin{proof}(Proof of Prop.~\ref{prop:bernstein-trace})
We have already established
\begin{align}
\langle W\rangle = \Sigma,
\end{align}
so $\langle\Tr W\rangle = 1$.
It remains only to bound the variation of $\Tr W$ away from this mean.

Since $W \sim \CWE(\Sigma, K)$,
\begin{align}
W = \frac{1}{K}\Sigma^{1/2} Y Y^\dagger \Sigma^{1/2}\;,
\end{align}
for $Y \in \C^{d\times K}$ with i.i.d. elements $Y_{ij} \sim \CN(0,1)$.
Work in an eigenbasis of $\Sigma$, and write $s_i$ for its eigenvalues. Then
\begin{align}
\begin{split}
\Tr W &= \frac{1}{K}\sum_{ij} s_i |Y_{ij}|^2 \\
&= \frac{1}{K}\sum_{ij} s_i \left[\left(\mathrm{Re}\;Y_{ij}\right)^2 + \left(\mathrm{Im}\;Y_{ij}\right)^2\right]\;.
\end{split}
\end{align}
The $\mathrm{Re},\;\mathrm{Im}\;Y_{ij}$ are Gaussian with variance $\frac 1 2$, so they have subgaussian norm
\begin{align}
\| \mathrm{Re},\;\mathrm{Im}\;Y_{ij} \|_{\psi_2} = \sqrt{4/3}
\end{align}
and their squares are subexponential%
\footnote{
By ``subexponential'' we mean \textit{light}-tailed---with exponential tail bounds, in analogy with ``subgaussian''.
This is in contrast with the family of heavy-tailed probability distributions.
}
with mean
\begin{align}
\langle(\mathrm{Re}\; Y_{ij})^2\rangle = \langle(\mathrm{Im}\;Y_{ij})^2\rangle = \frac 1 2 
\end{align}
and subexponential norm
\begin{align}
E = 4/3\;.
\end{align}
The normalization then has the form 
\begin{align}
\Tr W = \frac{1}{2K}\sum b_{\alpha} + \frac{1}{K}\sum b_\alpha (Z_\alpha - 1/2)
\end{align}
where the $b_\alpha$ consist of $2K$ copies of each eigenvalue of $\Sigma$ and the $Z_\alpha$ are $2dK$ independent real random variables (the squared real and imaginary parts) with subexponential norm $4/3$ and mean $1/2$.
In particular,
\begin{align}
\begin{split}
\|b\|_2^2 &= 2K \sum_j s_j^2 = 2K \|\Sigma\|^2_2 \\
\|b\|_\infty &= \max s_j = \|\Sigma\|_{\mathrm{op}}\;.
\end{split}
\end{align}
The subexponential Bernstein inequality~\cite[Thm.~2.9.1, Cor.~2.9.2]{HighDimensionalProbability}
then gives 
\begin{align}
\begin{split}
P\Big(|\Tr W - \langle \Tr W\rangle|\ge t\Big)
&\le 2 \exp\left[- c \min\left(\frac{K^2t^2}{E^2\|b\|_2^2}, \frac {K t} {E \|b\|_\infty}\right)\right] \\
&= 2 \exp\left[- c \min\left(\frac{9K^2t^2}{32K\|\Sigma\|^2_2}, \frac {3Kt} {4 \|\Sigma\|_{\mathrm{op}} }\right)\right]\;.
\end{split}
\end{align}
The corollary follows from
\begin{align}
\begin{split}
\|\Sigma\|_2^2 &= \Tr[\rho_0^2] \\
\|\Sigma\|_{\mathrm{op}} &=\|\rho_0\|_{\mathrm{op}}\; .
\end{split}
\end{align}
\end{proof}

\subsection{Empirical spectral distributions (ESDs) of ME, FTCWE and CWE}\label{app:esd-ftcwe-cwe}

Here we consider the relationship of the spectra of FTCWE and CWE matrices.
First we show that they are close with high probability---this follows almost immediately from concentration of trace.
Then we consider rescaled empirical spectral distributions, which are the objects of interest in free probability;
finally we comment on the limitations of our results.

\begin{prop}
Take $W \sim \CWE(\Sigma, K)$, $\Sigma = \rho_0$ a reference density matrix, and $\rho = W/\Tr W$, so $\rho \sim \FTCWE(\Sigma)$.
Write 
\begin{align}
\sigma^\dn(W), \sigma^\dn(\rho)
\end{align}
for the vectors of eigenvalues of $W$ and $\rho$ in non-increasing order.
Then the distance between spectra is small, even compared to the magnitudes of the spectra:
\begin{align}
P\left(\frac{\|\sigma^\dn(W) - \sigma^\dn(\rho)\|_2}{\|\sigma^\dn(\rho)\|_2} > t \right) \le
2 \exp\left[- \kappa \min\left(\frac{9Kt^2}{32\Tr[\rho_0^2]}, \frac {3Kt} {4 \|\rho_0\|_{\mathrm{op}} }\right)\right]\;.
\end{align}
\end{prop}

\begin{proof}
The distance between spectra is
\begin{align}
\begin{split}
\|\sigma^\dn(W) - \sigma^\dn(\rho)\|_2^2 &= \|\sigma^\dn(\rho)(\Tr W -1) \|_2^2 \\
&= \|\sigma^\dn(\rho)\|_2^2 |\Tr W - 1|^2 \;.
\end{split}
\end{align}
The result follows immediately from concentration of normalization, Cor.~\ref{cor:trace-concentration}.
\end{proof}
So far in this section we have shown that CWE spectra are close to FTCWE spectra.
In Sec.~\ref{sss:integrability} we showed that the FTCWE is close to the ME---our real object of interest---as far as Lipschitz functions, and by the Hoffman-Wielandt inequality, the map from matrix to spectrum is Lipschitz:
\begin{align}
\|\sigma^\dn(A) - \sigma^\dn(B)\|_2 \le \|A - B\|_2\;.
\end{align}
So for ``reasonable''---Lipschitz, with coefficient uniform in $d$---functions of the spectrum $\sigma(\rho)$, we can replace ME by CWE without trouble.

But in free probability it is necessary to rescale the matrices by a factor of $d$,
so that the empirical spectral distributions can converge as $d, K \to \infty$.
(One way to see this is that the average of the eigenvalues of a density matrix on $\C^d$ is immediately $1/d$.)
The natural object is in fact
\begin{align}
\mu_{dA} = \frac 1 d \sum_j \delta_{da_j}\;,
\end{align}
where $a_j$ are the eigenvalues of the matrix $A$. 
In the following Propositions and Corollary, we show that the ME, FTCWE, and POVM ensembles give empirical spectral measures that are close to each other and to the generalized microcanonical ensemble.
\begin{prop}
Write $d$-point empirical measures
\begin{align}
    \mu_x = \frac 1 d \sum_j \delta_{x_j}\;,
\end{align}
$x \in \R^d$ and $\delta_{x_j}$ the Dirac delta function (atom) at $x_j$;
write $\mathcal E_d$ for the set of such measures
\begin{align}
\mathcal E_d = \left\{\frac 1 d \sum_{j = 1}^d \delta_{x_j}\;, x_j \in \R\right\}
\end{align}
and take the standard metric on $\R$ $D(x,y) = |x-y|$.
The Wasserstein $p$-distance between $d$-point empirical measures with respect to that metric is bounded
\begin{align}
 W_p(\mu_x, \mu_y) \le d^{-1/p} \|x^\dn - y^\dn\|_p\;,
\end{align}
where $x^\dn$ is the vector $x$ permuted into non-increasing order.
\end{prop}
\begin{proof}
Consider the coupling $\Gamma = \{(x^\dn_j, y^\dn_j)\}$. The expectation value of the distance in this coupling gives
\begin{align}
\begin{split}
W_p(\mu_x, \mu_y) &\le \mathds E_{(x,y) \sim \Gamma} [D(x,y)^p ]^{1/p} \\
&= d^{-1/p}\left[\sum_j |x^\dn_j - y^\dn_j|^p \right]^{1/p} \\
&= d^{-1/p} \|x^\dn - y^\dn\|_p\;.
\end{split}
\end{align}
\end{proof}

\begin{prop}
Take $\mathcal E_d$ to be a metric space with metric given by the Wasserstein $p$-distance, $p \ge 1$.
The map taking sphere variables to rescaled empirical measures
\begin{align}
\begin{split}
\mathrm{spec} &: S^{2dK-1} \to \mathcal E_d \\
\mathrm{spec} &: \omega \mapsto \mu_{d\omega\omega^\dagger}
\end{split}
\end{align}
has Lipschitz coefficient
\begin{align}
L \le 2d^{1 - 1/p}\;.
\end{align}
\end{prop}
\begin{proof}
Take $\omega, \varphi \in S^{2dK-1}$ and write unrescaled spectra
\begin{align}
\begin{split}
x &= \mathrm{spec}\; \omega\omega^\dagger \\
y &= \mathrm{spec}\; \varphi\varphi^\dagger \\
\end{split}
\end{align}
where the spectra are in non-increasing order.
For $p \ge 1,$
\begin{align}
\begin{split}
W_p(\mu_{dx}, \mu_{dy}) 
&\le d^{1-1/p} \|x - y\|_p \\
&\le d^{1-1/p} \|x - y\|_1 \\
&\le d^{1-1/p}\|\omega\omega^\dagger - \varphi\varphi^\dagger\|_1 \\
&\le 2d^{1-1/p}\|\omega (\omega - \varphi)^\dagger\|_1 \\
&\le 2d^{1-1/p}\|\omega\|_2 \|\omega - \varphi\|_2 \\
&\le 2d^{1-1/p}\|\omega - \varphi\|_2
\end{split}
\end{align}
using monotonicity of $\ell_p$ norms, H\"older's inequality for Schatten norms, and the sphere constraint $\|\omega\|_2 = 1$.
\end{proof}
\begin{rem}
This establishes that the rescaled ESDs of the ME, FTCWE, and POVM ensembles are close to each other and to the generalized microcanonical ensemble, in the sense that certain functions of the rescaled spectra have expectation values that approach each other.
For Lipschitz functions, this comes from the $p = 1$ case via Kantorovich-Rubinstein and the error is
\begin{align}
\propto 1/\sqrt K\;.
\end{align}

But many functions of interest, e.g. entropies, are neither moments nor Lipschitz, so these arguments do not directly apply.
We expect that a truncation argument can be used to show that those functions are close in the three ensembles,
but a slightly different argument would be required for each function.
We further expect that an appropriate notion of distance between the ESD measures would control those entropy errors,
but we do not know what that notion of distance is.
\end{rem}
\begin{rem}
We have shown that spectra of the FTCWE are close to those of the generalized microcanonical ensemble.
One might worry that spectra of the generalized microcanonical ensemble are precisely those of the reference state---but this turns out not to be the case.
If the reference state is
\begin{align}
\Sigma_0 = \sum_{j = 1}^d s_j \ketbra{j}{j}\;,
\end{align}
a state from the generalized microcanonical ensemble can be written in terms of a sphere variable
\begin{align}
\Omega = \sum_{j = 1}^d \sqrt{s_j} \ket{j}\otimes \ket{\Theta_j}
\end{align}
where the $\ket{\Theta_j}$ are i.i.d. Haar on $\C^K$, resulting in a density matrix 
\begin{align}
\rho = \sum_{j,k = 1}^d \Big(\sqrt{s_j s_k} \braket{\Theta_j}{\Theta_k}\Big) \ketbra{j}{k}\;.
\end{align}
The $\ket{\Theta_j}$ are not orthogonal: they are i.i.d. Haar, so the off-diagonal entries in the Gram matrix is $\braket{\Theta_j}{\Theta_k} \sim 1/\sqrt K$ and the spectrum is not simply $s_j$.

\end{rem}

\section{Derivation of S-transform expression}
\label{app:S-transform}
Consider a Hamiltonian $H$ with spectral density $\sigma(E)$ supported on $[E_-, E_+]$ (where either may be infinite). We assume $\beta>0$ throughout. In the large-$d$ limit, the partition function satisfies
\begin{equation}
\label{UbetaGeneral}
    \lim_{d\to \infty}\frac{Z}{d}= \int_{E_-}^{E_+} \sigma(E) \, e^{-\beta E} \, dE \coloneq  U(\beta)\; ,
\end{equation}
where it is assumed that $0<U(\beta)<\infty$. The rescaled eigenvalue of the thermal state is therefore given by
\begin{equation}
    x = \frac{d \, e^{-\beta E}}{Z}= \frac{e^{-\beta E}}{U(\beta)}, \quad x\in[x_-, x_+],
\end{equation}
where $x_\pm = \frac{e^{-\beta E_{\mp}}}{U(\beta)}$. We transform $\sigma(E)$ to a density in terms of the rescaled eigenvalue $x$ via
\begin{align}
    \pi_{\rho_0}(x) &= \sigma\left[E(x)\right]\abs{\frac{dE}{dx}}\nonumber\\
    &=\frac{1}{\beta x} \sigma\left[E(x)\right]
\end{align}
where we have used the fact that the derivative $\abs{\frac{dE}{dx}}$ of $E(x) = -\beta^{-1}\log\left[ x U(\beta)\right]$ is $\frac{1}{\beta x}$. The S-transform of the above thermal density is then computed using the standard three-step procedure~\cite{Voiculescu}:
\begin{enumerate}
    \item compute the moment generating series $\psi_{\rho_0}(z) = \sum_{n=1}^\infty m_n z^n$, where $m_n = \int_{x_-}^{x_+}dx \ \pi_{\rho_0}(x) x^n$ are the moments of $\pi_{\rho_0}(x)$. 
    \item find the inverse under composition of $\psi_{\rho_0}$, i.e. the function $\chi_{\rho_0}$ such that $\chi_{\rho_0}\left(\psi_{\rho_0}(z)\right)=z$.
    \item the S-transform is then
    \begin{align}
    \label{SChiRelation}
        S_{\rho_0}(z) = \frac{1+z}{z}\chi_{\rho_0}(z) . 
    \end{align}
\end{enumerate}
First, we express the $n^{\text{th}}$ moment
\begin{equation}
    m_n = \int_{x_-}^{x_+}dx \ \pi_{\rho_0}(x) \ x^n = \int_{E_-}^{E_+} dE \ \sigma(E) \left(\frac{e^{-\beta E}}{U(\beta)}\right)^n= \frac{U(n\beta)}{\left(U(\beta)\right)^n}.
\end{equation}
Substituting this in, we obtain the series
\begin{align}
    \psi_{\rho_0}(z) &= \sum_{n=1}^\infty \frac{U(n\beta)}{\left(U(\beta)\right)^n} \, z^n\nonumber\\
    &=  \sum_{n=1}^\infty \left(\frac{z}{U(\beta)}\right)^n \int_{E_-}^{E_+} dE \ e^{-n\beta E}\sigma(E)\nonumber\\
    &=\int_{E_-}^{E_+} dE \ \sigma(E)\left( \sum_{n=1}^\infty \left(\frac{z e^{-\beta E}}{U(\beta)}\right)^n\right) \nonumber\\
    &= \int_{E_-}^{E_+} dE \ \sigma(E) \ \frac{\frac{ze^{-\beta E}}{U(\beta)}}{1-\frac{ze^{-\beta E}}{U(\beta)}}\nonumber\\
    &= z\int_{E_-}^{E_+} dE \ \frac{\sigma(E)}{U(\beta)e^{\beta E}-z}.
\end{align}
Then, the inverse function $\chi_{\rho_0}$ is the solution of the implicit equation
\begin{equation}
    z = \chi_{\rho_0}(z)\int_{E_-}^{E_+} dE \ \frac{\sigma(E)}{U(\beta)e^{\beta E}-\chi_{\rho_0}(z)},
\end{equation}
which, using the relation~\eqref{SChiRelation} alongside the S-transform $S_c(z) = (1+cz)^{-1}$ of the MP distribution, results in the S-domain function
\begin{equation}
\label{STransformGeneral}
    S(z) = S_c(z) S_{\rho_0}(z)= \frac{(1+z)}{1+cz}\left[\int_{E_-}^{E_+} dE \ \frac{\sigma(E)}{U(\beta)e^{\beta E}-\frac{z(1+cz)}{1+z}S(z)}\right]^{-1}.
\end{equation}
The inverse S-transform is implemented in our accompanying numerics~\cite{Code-For-Figures-Supp} using a change of variables. Let $G(x)$ be the Cauchy transform of the eigenvalue density $\pi(x)$. Since $G(x)=\frac{1}{x}\left[1+\psi\left(\frac{1}{x}\right)\right]$, the complex variable $z=\psi(1/x)$, which satisfies $\chi(z)=1/x$, is related to the (real) physical eigenvalue $x$ by
\begin{equation}
\label{zlambda}
    z = x G(x)-1,
\end{equation}
Hence,
\begin{equation}
\label{SGlambda}
    S(z) = \frac{1+z}{z}\chi(z) = \frac{G(x)}{x G(x)-1}.
\end{equation}
Now, define the auxiliary variable
\begin{equation}
    u = \frac{z(1+cz)}{1+z}S(z),
\end{equation}
and substitute the pair~\eqref{zlambda} and~\eqref{SGlambda} into the definition of $u$. This yields
\begin{equation}
    u = \frac{1+c(x G(x)-1)}{x}.
\end{equation}
In terms of $u$, the integral~\eqref{STransformGeneral} becomes
\begin{equation}
    \int \frac{\sigma(E) dE}{U(\beta)e^{\beta E}-u}\coloneq  H(u),
\end{equation}
so the implicit equation becomes
\begin{equation}
    S(z) = \frac{1+z}{1+cz}\frac{1}{H(u)}.
\end{equation}
Substituting in the pair~\eqref{zlambda} and~\eqref{SGlambda} again, we obtain the relation
\begin{equation}
\label{lambda-H(u)}
    x = \frac{1}{u}+cH(u),
\end{equation}
from which we compute the root $u$ corresponding to a given eigenvalue $x$. It remains to reconstruct the Cauchy transform $G(x)$, and then to obtain from it the density $\pi(x)$. From $x G(x)=1+uH(u)$ alongside~\eqref{lambda-H(u)}, we arrive at
\begin{equation}
    G(x) = \frac{u(1+u H(u))}{1+cu H(u)}.
\end{equation}
Finally, using the \textit{Sokhotski-Plemelj} theorem $\pi_{\text{cont}}(x) =- \frac{1}{\pi}\Im{G(x+i0)}$~\cite{RandomMatricesTheoryPractice}, we obtain the continuous part of the density numerically.
\subsection{Gaussian energy density}
\label{app:Gaussian-spec-dens}
For a Hamiltonian with the Gaussian spectral density
\begin{equation}
\label{GaussianSpecDensSupplementary}
    \sigma(E)= \frac{1}{\sigma_0\sqrt{2\pi}}\, e^{-E^2/2\sigma_0^2}\; ,
\end{equation}
the two-sided Laplace transform~\eqref{UbetaGeneral} is
\begin{equation}
    U(\beta) = e^{\beta^2\sigma_0^2/2}\;.
\end{equation}
With this and the S-transform relation~\eqref{STransformGeneral}, we may compute the eigenvalue density at arbitrary temperature (see Fig.~\ref{CWELevelDensity} in the main text).

At high temperatures $\beta\sigma_0 \ll 1$, $U(\beta) \approx 1+\frac{\beta^2\sigma_0^2}{2}$ to leading order. The $n^{\text{th}}$ moment then becomes
\begin{align}
    m_n \approx \frac{1+\frac{n^2\beta^2\sigma_0^2}{2}}{\left(1+\frac{\beta^2\sigma_0^2}{2}\right)^n} \approx \left(1+\frac{n^2\beta^2\sigma_0^2}{2}\right)\left(1-\frac{n\beta^2\sigma_0^2}{2}\right)= 1+\frac{\beta^2\sigma_0^2}{2}n(n-1) + \mathcal{O}(\beta^4)\;,
\end{align}
where we have used the binomial approximation $(1+x)^{-n}\approx 1-nx$ to go from the first to the second equality. This gives the $\psi$ series
\begin{align}
    \psi(z)&\approx \sum_{n=1}^\infty z^n + \frac{\beta^2\sigma_0^2}{2}\sum_{n=1}^\infty n(n-1)z^n\nonumber\\
    &= \sum_{n=1}^\infty z^n + \frac{\beta^2\sigma_0^2}{2}z^2\sum_{n=2}^\infty n(n-1)z^{n-2}\nonumber\\
    &= \frac{z}{1-z}+\frac{\beta^2\sigma_0^2z^2}{(1-z)^3}.
\end{align}
We now require the inverse under composition $\chi$. For $\beta = 0$ this is $\chi(z) = \frac{z}{1+z}$. Given this, we propose the perturbed Ansatz:
\begin{equation}
    \chi(z)= \frac{z}{1+z}+\beta^2\sigma_0^2f(z).
\end{equation}
Substituting this back into $\psi(\chi(z))=z$ and discarding terms beyond $\mathcal{O}(\beta^2)$, we find $f(z) = -\frac{z^2}{1+z}$. This gives
\begin{equation}
    \chi_{\rho_0}(z)= \frac{z}{1+z}-\beta^2\sigma_0^2\frac{z^2}{1+z}= \frac{z(1-\beta^2\sigma_0^2z)}{1+z}, 
\end{equation}
meaning the S-transform is the linear expression
\begin{equation}
    S_{\rho_0}(z) = 1-\beta^2   \sigma_0^2z ,
\end{equation}
and the overall $S$-domain function is simply
\begin{equation}
\label{STransformHighTemp}
    S(z) = \frac{1-\beta^2\sigma_0^2 z}{1+cz}. 
\end{equation}
We now work backwards to invert~\eqref{STransformHighTemp}. Using the relations between $S, \chi$ and $\psi$, alongside the definition of the Cauchy transform $G(\lambda) = \frac{1}{\lambda}\left(1+\psi\left(\frac{1}{\lambda}\right)\right)$, we set $w = \psi\left(\frac{1}{\lambda}\right)$ and solve $\chi(w) = 1/\lambda$, i.e.
\begin{equation}
    \frac{1}{\lambda}= \frac{w(1-\beta^2\sigma_0^2w)}{(1+w)(1+cw)}\;.
\end{equation}
Rearranging this into a quadratic expression in $w$, we write the solution
\begin{equation}
    w = \frac{-(1+c-\lambda)\pm\sqrt{(1+c-\lambda)^2-4(c+\lambda \beta^2\sigma_0^2)}}{2(c+\lambda\beta^2\sigma_0^2)}.
\end{equation}
Finally, taking the branch of the square root with $\Im\,w(x+i0)<0$ (so that $\pi_{\text{cont}}\geq0$), we obtain the continuous part of the level density from the imaginary part of the Cauchy transform:
\begin{align}
    \pi_{\text{cont}}(x) &= -\frac{1}{\pi}\Im{G(x+i0)} \nonumber\\
    &= -\frac{1}{\pi x}\Im{w(x+i0)}\nonumber\\
    &= \frac{\sqrt{4(c+\beta^2\sigma_0^2 x)-(1+c-x)^2}}{2\pi x (c+\beta^2\sigma_0^2 x)}.
\end{align}
\subsection{Random (GUE) Hamiltonian}
\label{app:GUE}
\begin{figure}[h]
    \centering
    \begin{subfigure}{0.45\linewidth}
        \centering
        \includegraphics[width=\linewidth]{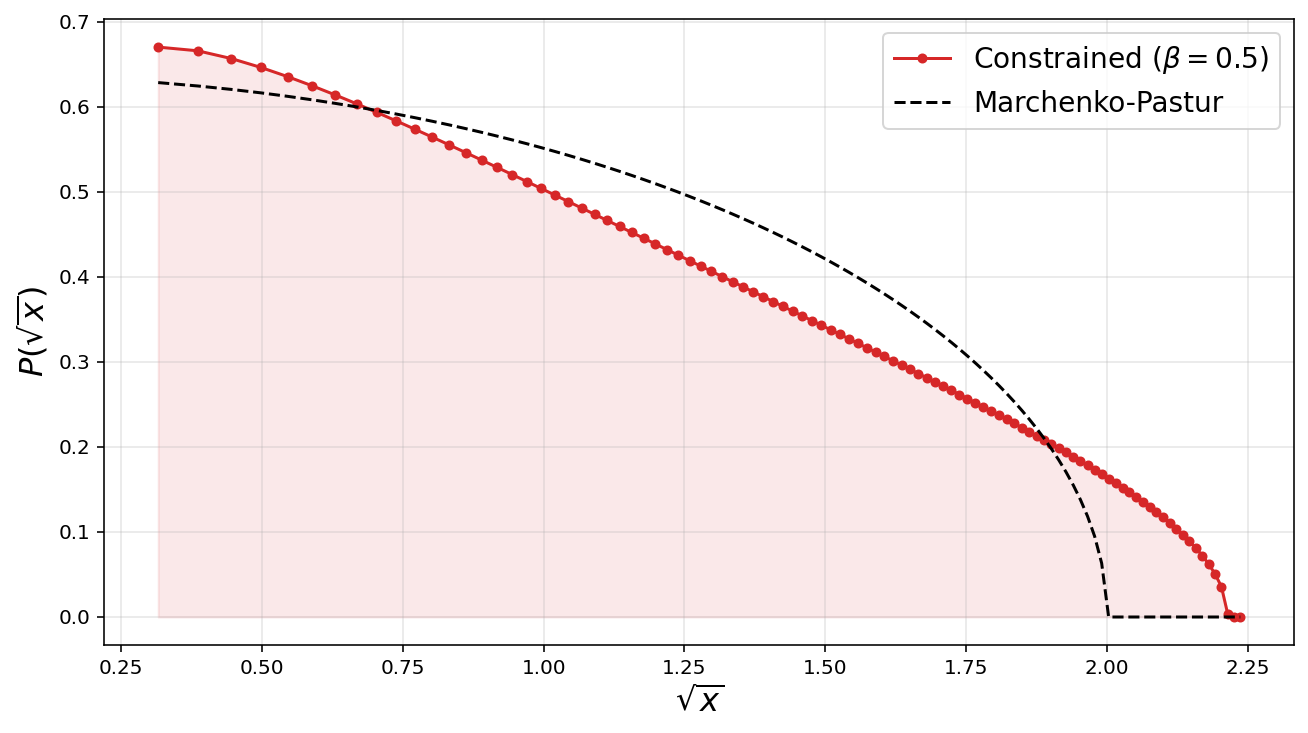}
    \end{subfigure}
    \hfill
    \begin{subfigure}{0.45\linewidth}
        \centering
        \includegraphics[width=\linewidth]{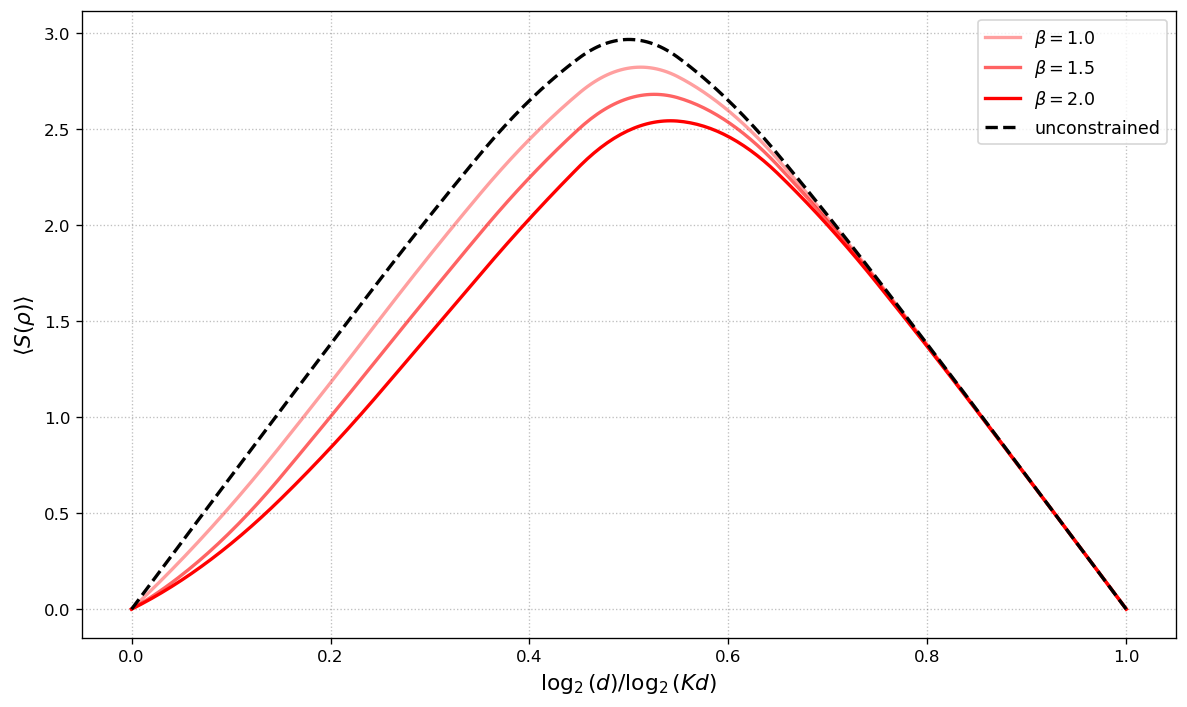}
    \end{subfigure}
    \caption{ \small (a) Level density of the thermally constrained ensemble in terms of $\sqrt{x}$ for $\beta = 0.5$, $c = 1$. At $\beta = 0$, the states are distributed according to the Hilbert-Schmidt measure with a quarter-circular density in $\sqrt{x}$. (b) Page curves under the thermal constraint for the variable $S(\rho)$ averaged first over the FTCWE and then over the GUE.}
    \label{GUESpectralPlots}
\end{figure}

Here we include the additional example of $\rho_0$ fixed to the thermal state of a random Hamiltonian. The natural unitarily invariant distribution over random Hermitian matrices is the Gaussian unitary ensemble (GUE), whose spectral density is given by the semicircular law~\cite{RandomMatricesTheoryPractice,RMTEdelmanRao,SpeicherRandomMatrices}
\begin{equation}
\label{GUESpecDens}
    \sigma(E) = \frac{1}{2\pi}\sqrt{4-E^2}\; , \quad E \in [-2,2].
\end{equation}
The two-sided Laplace transform is then
\begin{equation}
\label{UbetaGUE}
    U(\beta) = \frac{I_1(2\beta)}{\beta}\;,
\end{equation}
where $I_1(\cdot)$ is the first order modified Bessel function. We substitute these into Eq.~\eqref{STransformGeneral} and invert the S-transform numerically to compute the level density at a given temperature $\beta^{-1}$ and aspect ratio $c$ (see Fig.~\ref{GUESpectralPlots}).

\begin{figure}[h]
    \centering
    \begin{subfigure}{0.45\linewidth}
        \centering
        \includegraphics[width=\linewidth]{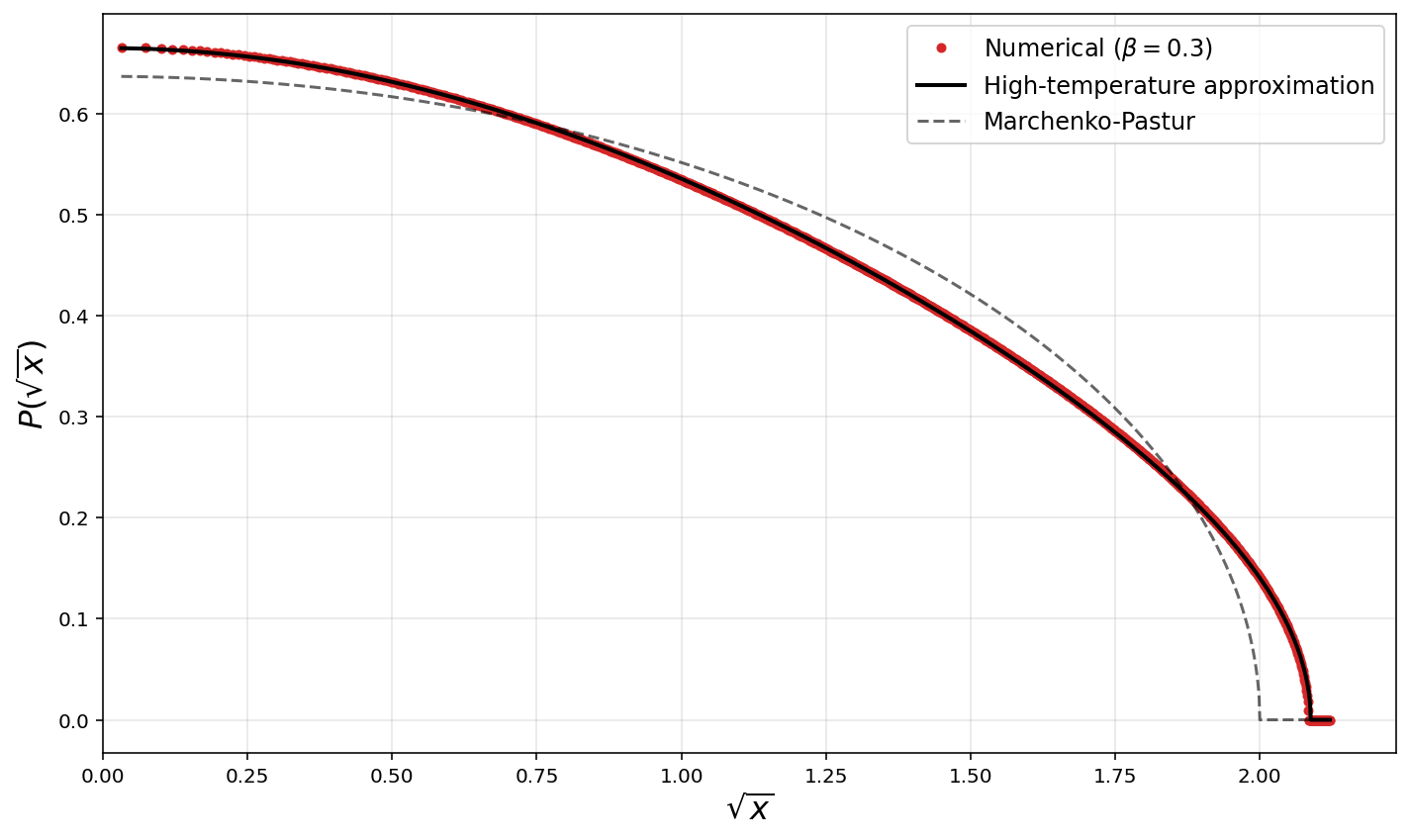}
        \caption{ \small }
    \end{subfigure}
    \hfill
    \begin{subfigure}{0.45\linewidth}
        \centering
        \includegraphics[width=\linewidth]{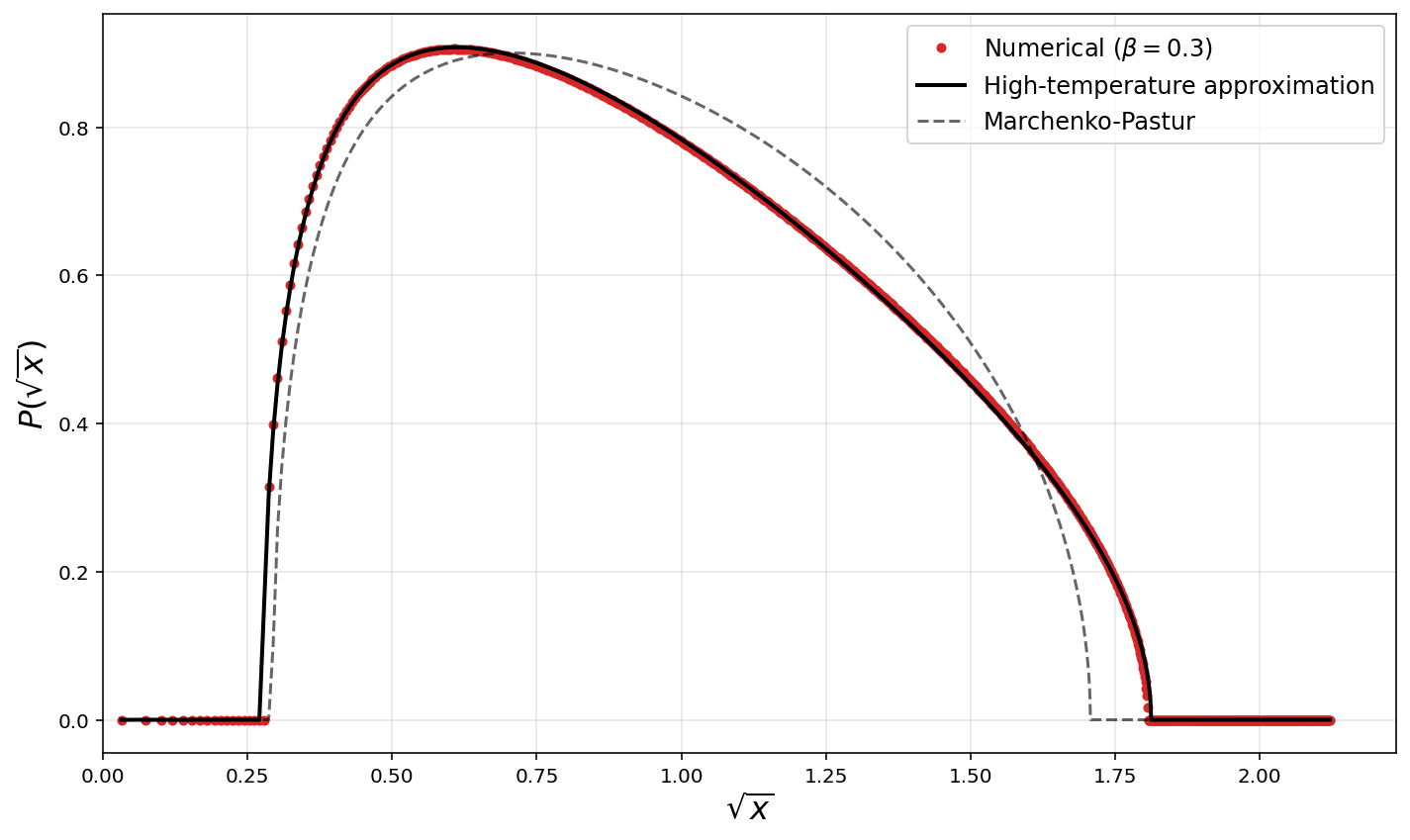}
        \caption{ \small }
    \end{subfigure}
    \caption{ \small Numerical simulations for the single-point densities of the rescaled eigenvalue $\sqrt{x}$ compared to the high-temperature approximation for (a) $\beta = 0.3$, $c = 1$, (b) $\beta = 0.3$, $c = 0.5$.}
    \label{HighTempGUE}
\end{figure}

At high temperatures $\beta \ll 1$, we approximate~\eqref{UbetaGUE} by
\begin{equation}
\label{UBetaGUEapprox}
    U(\beta) \approx 1+\frac{\beta^2}{2}\; . 
\end{equation}
The $n^{\text{th}}$ moment is then $m_n\approx 1+\frac{\beta^2}{2}n(n-1)$ to leading order, giving the moment generating series
\begin{equation}
    \psi(z) = \frac{z(1-z)^2+\beta^2 z^2}{(1-z)^3}. 
\end{equation}
Applying the same perturbed ansatz as in the Gaussian case (but with $\sigma_0=1$), we obtain the inverse under composition
\begin{equation}
    \chi(z) = \frac{z(1-\beta^2 z)}{1+z}\;;
\end{equation}
the corrected continuous part of the MP law~\eqref{MarchenkoPastur} is then
\begin{equation}
    \pi_\text{cont}(x)\approx\frac{\sqrt{4(c+\beta^2 x)-(1+c-x)^2}}{2\pi x(c+\beta^2 x)}.
\end{equation}
Fig.~\ref{HighTempGUE} shows the agreement between this high-temperature expression and the density obtained by numerical inversion of $S(z)$. 
\section{Random POVM}
\label{app:random-POVM}
Consider a bipartite system with $\mathcal{H}_A = \mathcal{H}_B \cong \mathbb{C}^d$ prepared in the canonical purification of $\rho_0 \coloneq  \sum_{i=1}^d\lambda_i\ketbra{i}{i}$, i.e.
\begin{equation}
    \ket{\psi}_{AB} = \sum_{i=1}^d \sqrt{\lambda_i}\ket{i}_A \otimes \ket{i}_B.
\end{equation}
From the $d\times K$ Ginibre matrix $X$, we construct a continuously indexed Kraus operator $M_X = K^{-1/2}X^T$. We act with this operation on the copy space $B$; this maps $\mathcal{H}_B\to\mathcal{H}_C\cong \mathbb{C}^K$, yielding the state
\begin{equation}
    \tilde{\ket{\psi}}_{AC}= \sum_{i=1}^d \sqrt{\lambda_i} \ket{i}_A \otimes M_X \ket{i}_B
\end{equation}
with probability density $p_X=\bra{\psi}(I_d \otimes M_X^\dagger M_X)\ket{\psi}$ with respect to the Gaussian measure $d\mu(X) = \pi^{-Kd}\, e^{-\Tr XX^\dagger} dX$. Tracing out the resulting $K$-dimensional ancilla $C$, the unnormalised reduced state of subsystem $A$ is
\begin{align}
\label{WXderivation}
    W_X &= \Tr_C \tilde{\ket{\psi}}\tilde{\bra{\psi}}\nonumber\\
    &= \Tr_C\left[\sum_{i=1}^d\sum_{j=1}^d\sqrt{\lambda_i}\sqrt{\lambda_j} \ketbra{i}{j} \otimes \left(M_X \ketbra{i}{j}M_X^\dagger\right) \right]\nonumber\\
    &=\sum_{i=1}^d \sum_{j=1}^d \sqrt{\lambda_i}\sqrt{\lambda_j} \ketbra{i}{j} \cdot \bra{j}M_X^\dagger M_X \ket{i}.
\end{align}
Evaluating the matrix element
\begin{equation}
    \bra{j}M_X^\dagger M_X \ket{i}= \frac{1}{K}\left(X^*X^T\right)_{ji}= \frac{1}{K}\sum_{k=1}^K X^*_{jk}X^T_{ki}= \frac{1}{K}\sum_{k=1}^K X^*_{jk}X_{ik}, 
\end{equation}
we compare it to the $ij^{\text{th}}$ element of the uncorrelated Wishart matrix $XX^\dagger$
\begin{equation}
    \left(XX^\dagger\right)_{ij}=\sum_{k=1}^K X_{ik}X^\dagger_{kj}=\sum_{k=1}^K X_{ik}X^*_{jk}
\end{equation}
to obtain the relation:
\begin{equation}
    \bra{j}M_X^\dagger M_X \ket{i} = \frac{1}{K}\left(XX^\dagger\right)_{ij}.
\end{equation}
Substituting this into~\eqref{WXderivation}, we rewrite the unnormalised reduced state as
\begin{equation}
    W_X = \frac{1}{K}\sum_{i=1}^d\sum_{j=1}^d \sqrt{\lambda_i}\sqrt{\lambda_j} \left(XX^\dagger\right)_{ij}\ketbra{i}{j}. 
\end{equation}
Finally, we insert the spectral decomposition of $\rho_0$ to express $W_X$ as the correlated Wishart matrix 
\begin{equation}
    W_X = \frac{1}{K}\sqrt{\rho_0}XX^\dagger\sqrt{\rho_0}\;.
\end{equation}
To verify that the effects $\{M_X^\dagger M_X\}$ form a POVM, we use that 
\begin{align}
    \langle M_X^\dagger M_X\rangle &= \frac{1}{K}\langle X^*X^T\rangle\nonumber\\
    &= \frac{1}{K}\langle(XX^\dagger)^*\rangle\nonumber\\
    &= \frac{1}{K}\langle XX^\dagger \rangle^*\nonumber\\
    &= I_d ,
\end{align}
which follows from the property $\langle XX^\dagger \rangle = K I_d$ of the Wishart ensemble.

If the $X$ were simply drawn from the unweighted Gaussian measure $d\mu(X)$, the physical states $\rho_X = W_X / \Tr W_X$ would be distributed according to the FTCWE with density~\eqref{FTCWEDensity}. However, our measurement protocol yields the state $\rho_X$ with Born-rule weighting $p_X = \Tr W_X$, i.e. the ensemble $\mathcal{E} = \{p_X, \rho_X \}$. The expectation value of an arbitrary test function $F(\rho)$ with respect to this ensemble is given by 
\begin{equation}
    \langle F(\rho)\rangle_\mathcal{E} = \int d\mu(X)\, p_X F(\rho_X),
\end{equation}
and this additional Born weighting means the ensemble has the density 
\begin{equation}
    P_\mathcal{E}(\rho)\propto \frac{\det(\rho)^{K-d}}{\left(\Tr \rho_0^{-1}\rho\right)^{dK+1}}\, \Theta(\rho)\delta(\Tr\rho-1).
\end{equation}
\begin{figure}[h]
    \centering
    \includegraphics[width=0.45\linewidth]{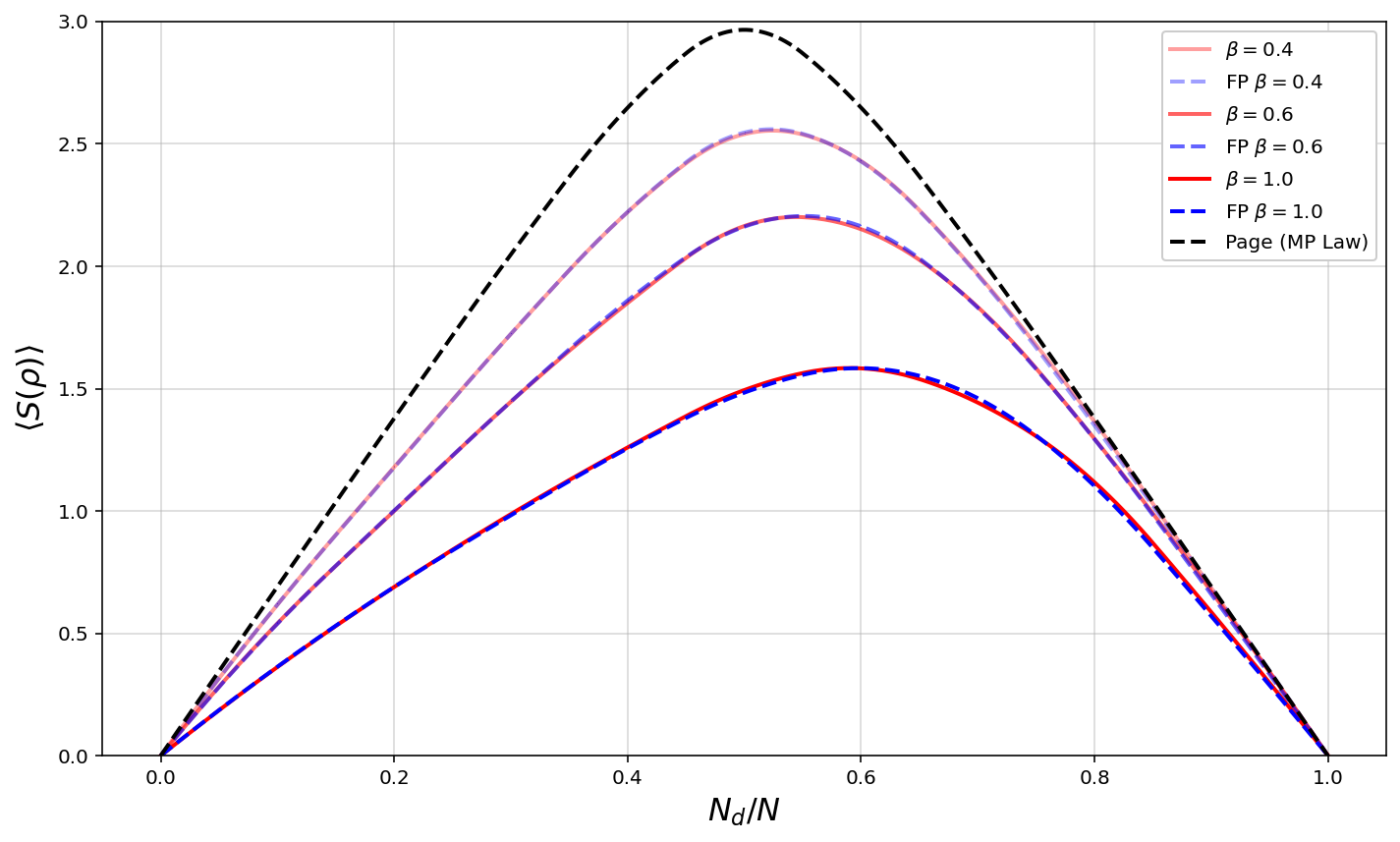}
    \caption{ \small Page curves for the POVM-generated ensemble $\mathcal{E}$ obtained by direct sampling (red) compared to the asymptotic free probability curves (blue, dotted). We set $\rho_0$ to the thermal state of the $N_d$-qubit TFIM ($J=h=1$), and fix $N=N_d+N_K=10$ in the direct (finite-size) sampling, with $K=2^{N_K}$.}
    \label{Page-curve-POVM-vs-free-prob}
\end{figure}

Fig.~\ref{Page-curve-POVM-vs-free-prob} verifies that the Page curves obtained by sampling from the ensemble $\mathcal{E}$ directly agree with the asymptotic predictions given by free probability for our TFIM example. The latter curves are generated by first inverting the S-domain relation~\eqref{STransformGeneral} numerically to obtain the level density $\pi(x)$, and then computing the integral
\begin{equation}
S_{\text{free}} = \ln d - \int \pi(x)\, x \ln x \, dx \; .
\end{equation}
\putbib[bibliography-CWE-supplemental]
\end{bibunit}
\end{document}